\documentclass[12pt]{article}

\usepackage{epsf}
\usepackage{a4}
\usepackage{multirow}
\usepackage{rotating}
\usepackage[tbtags]{amsmath}

\newcommand{\no}{\nonumber}
\newcommand{\q}{\quad}
\newcommand{\qq}{\qquad}

\newcommand{\overcirc}[1]{\overset{\circ}{#1}}

\begin{document}

\hfill 

\hfill 

\bigskip\bigskip

\begin{center}

{{\Large\bf  $\mbox{\boldmath$U(3)$}$ 
chiral perturbation theory \\ with infrared regularization
\footnote{Work supported in part by the DFG}}}

\end{center}

\vspace{.4in}

\begin{center}
{\large B. Borasoy\footnote{email: borasoy@physik.tu-muenchen.de}
and S. Wetzel\footnote{email: swetzel@physik.tu-muenchen.de}}

\bigskip

\bigskip

Physik Department\\
Technische Universit{\"a}t M{\"u}nchen\\
D-85747 Garching, Germany \\

\vspace{.2in}

\end{center}

\vspace{.7in}

\thispagestyle{empty} 

\begin{abstract}
We include the $\eta'$ in chiral perturbation theory without employing
1/$N_c$ counting rules. The method is illustrated by calculating the masses 
and decay constants of the Goldstone boson octet $(\pi, K, \eta)$ and the
singlet $\eta'$  up to one-loop order.
The effective Lagrangian describing the interactions of the $\eta'$ with the
Goldstone boson octet is presented up to fourth chiral order and the loop
integrals are evaluated using infrared regularization, which preserves Lorentz
and chiral symmetry.
\end{abstract}

\vfill

\section{Introduction}  
Chiral perturbation theory is the effective field theory of QCD
at low energies. The QCD Lagrangian with massless quarks
exhibits an $SU(3)_R \times SU(3)_L$ chiral symmetry which is broken down
spontaneously to $SU(3)_V$, 
giving rise to a Goldstone boson octet of pseudoscalar mesons:
pions, kaons and the $\eta$,
which become massless in the chiral limit of zero quark masses.
The axial $U(1)$ anomaly, on the other hand, prevents the corresponding singlet
state, the $\eta'$, from becoming massless even in the chiral limit. 
Therefore, in conventional chiral perturbation theory the $\eta'$ 
is not included explicitly, although it does show up in the form of a
contribution to a coupling coefficient of the Lagrangian, a so-called
low-energy constant (LEC).

In the large $N_c$ limit the quark loop which is responsible for the $U(1)$
anomaly is suppressed. The chiral symmetry of the QCD Lagrangian is thus
extended to $U(3)_R \times U(3)_L$. At the level of the effective theory, the
octet of Goldstone bosons converts into a nonet with $\eta'$ being the ninth
member. The properties of this particle are subject to similar constraints as
the original Goldstone bosons ($\pi, K, \eta$) and the corresponding effective
Lagrangian can be constructed. A systematic expansion of the Green functions of
chiral perturbation theory in powers of momenta, quark masses and $1/N_c$ was
introduced in \cite{GL} and has recently been more firmly established in
\cite{Leu, KL} and \cite{H-S}. 
An investigation of these papers reveals that in order to construct the
effective Lagrangian including the singlet field no large $N_c$ arguments are
necessary. An additional 1/$N_c$ counting scheme is imposed only to ensure that
loops with an $\eta'$ are suppressed by powers of 1/$N_c$.
In particular, the mass of the $\eta'$ which introduces an additional low
energy scale of about 1 GeV is proportional to 1/$N_c$ and can therefore be
treated perturbatively.

Without invoking large $N_c$ arguments the inclusion of the $\eta'$ in baryon
chiral perturbation theory has been outlined in \cite{B1}.
This work suffers, however, from the fact that at one loop order the $\eta'$
loop contributions are substantial due to the large mass of the $\eta'$, 
$m_{\eta'}$.
A systematic expansion in the meson masses is possible but its convergence is
doubtful, since $m_{\eta'}$ is close to the scale of chiral symmetry breaking
$\Lambda_\chi = 4 \pi F_\pi \sim 1.2$ GeV with $F_\pi \approx 93$ MeV the pion
decay constant.
This problem arises for every particle in the effective field theory which has
been included explicitly but has a mass similar to or larger than
$\Lambda_\chi$. 
The inclusion of the lowest lying baryon octet $(N,\Lambda,\Sigma,\Xi)$, e.g.,
spoils the chiral counting scheme if one employs the relativistic Lagrangian
and regularizes loop integrals dimensionally.
This can be prevented by going to the nonrelativistic limit and treating the
baryons as heavy static sources.
Within the so-called heavy baryon chiral perturbation theory one obtains a
one-to-one correspondence between the number of loops and the chiral order,
i.e. a chiral counting scheme emerges \cite{BKM}.
More recently an alternative way of treating massive fields has been proposed
in \cite{ET} and put on a more solid basis in \cite{BL}.
The authors keep the relativistic formulation of the Lagrangian but modify the
loop integrals in a chiral invariant way. In \cite{BL} this was achieved by
employing a modified regularization scheme, the so-called infrared
regularization, in which Lorentz and chiral invariance are kept at all stages.

The purpose of the present work is to implement infrared regularization in an
effective theory including the $\eta'$, while not using any large $N_c$
arguments. In this introductory presentation, we will restrict ourselves to the
purely mesonic case and calculate the masses and decay constants of the
pseudoscalar mesons.
We start in the following section by presenting the effective Lagrangian.
The dependence of the LECs on the renormalization scale in QCD is discussed in
Sec. 3 and App. A.
Section 4 is a presentation of our results for the masses and decay constants
up to one-loop order using infrared regularization.

\section{The effective Lagrangian}
The full effective $U(3)_R \times U(3)_L$ Lagrangian up to fourth chiral order,
i.e. including terms up to four derivatives and quadratic in the quark masses,
has already been given in \cite{H-S}.
We will therefore restrict ourselves to a repetition of the construction
principles which will make it obvious that no $1/N_c$ arguments are required.
To this end, consider the QCD Lagrangian in the presence of external sources
\begin{equation}  \label{lag}
{\cal L}_{\scriptscriptstyle QCD} = {\cal L}_{\scriptscriptstyle QCD}^0 + 
\bar{q} \gamma_\mu ( v^\mu + \gamma_5
 a^\mu) q -   \bar{q} ( s - i \gamma_5 p ) q 
  - \frac{g^2}{16 \pi^2} \theta(x) \mbox{tr}_c
  ( G_{\mu \nu} \tilde{G}^{\mu \nu} )
\end{equation}
with $\tilde{G}_{\mu \nu} = \epsilon_{\mu \nu \alpha \beta} G^{\alpha
\beta}$ and $\mbox{tr}_c$ is the trace over the color indices.
The term ${\cal L}_{\scriptscriptstyle QCD}^0$ 
describes the limit where the masses of the three
light quarks and the vacuum angle are set to zero and the external sources 
$v_\mu(x), a_\mu(x), s(x), p(x)$ are hermitian $3 \times 3$ matrices in flavor
space. The mass matrix of the three light quarks is contained in the external
field $s$.
Under $U(1)_R \times U(1)_L$ the axial $U(1)$ anomaly 
adds a term $ -( g^2 / 16 \pi^2)
2 N_f \, \alpha \, \mbox{tr}_c ( G_{\mu \nu} \tilde{G}^{\mu \nu} )$ to the
QCD Lagrangian, with $N_f$ being the number of different quark flavors and
$\alpha$ the angle of the axial $U(1)$ rotation.
The vacuum angle $\theta(x)$ is in this context treated as an external field
that transforms under an axial $U(1)$ rotation as
\begin{equation}
\theta(x) \rightarrow  \theta'(x) = \theta(x) - 2 N_f \alpha .
\end{equation}
Then the term generated by the anomaly in the fermion determinant is
compensated by the shift in the $\theta$ source and the Lagrangian from
Eq. (\ref{lag}) remains invariant under axial $U(1)$ transformations.
The original symmetry group $SU(3)_R \times SU(3)_L$ of the Lagrangian 
${\cal L}_{\scriptscriptstyle QCD}$ is extended to 
$G= U(3)_R \times U(3)_L$.\footnote{To be more
precise, the Lagrangian changes by a total derivative which gives rise to the
Wess-Zumino term. We will neglect this contribution since the corresponding
terms involve five or more meson fields which do not play any role for the
discussions here.}
This property remains at the level of an effective theory and the
additional source $\theta$ also shows up in the effective Lagrangian.
We assume that this extended symmetry $G$ is spontaneously broken down to $H=
U(3)_V$. The nine parameters of the coset space $G/H = U(3)$ correspond then to
the lowest lying nonet of pseudoscalar mesons: pions, kaons, $\eta$ and
$\eta'$. 
They can be most conveniently summarized in a matrix valued field
\begin{equation}
U(\phi,\psi) = u^2 (\phi,\psi) = 
\exp \lbrace  2 i \phi / F + i \psi /3\rbrace  ,
\end{equation}
where $F$ is the decay constant of the Goldstone boson octet $\phi$ 
in the chiral limit.
The unimodular part of the field $U(x)$ contains the degrees of freedom of the
octet $\phi$ 
\begin{eqnarray}
\phi =  \frac{1}{\sqrt{2}} 
\begin{pmatrix}
{\frac{1}{\sqrt 2}} \pi^0 + {\frac{1}{\sqrt 6}} \eta_8 &\pi^+ &K^+ \\
\pi^- & -{\frac{1}{\sqrt 2}} \pi^0 + {\frac{1}{\sqrt 6}} \eta_8 & K^0 \\
K^- & \bar{K^0}&- {\frac{2}{\sqrt 6}} \eta_8
\end{pmatrix} \, \, \, \, \, ,  
\end{eqnarray}
while the phase det$U(x)=e^{i\psi}$
describes the singlet $\eta_0$ .
The effective Lagrangian is formed with the fields $U(x)$,  
derivatives thereof and also includes the external fields: 
${\cal L}_{\mbox{\footnotesize eff}}(U,\partial U,\ldots, v, a, s, p,\theta)$. 
Under $U(3)_R \times U(3)_L$ the fields transform as
follows
\begin{eqnarray} \label{trafo}
U' &=& RUL^{\dagger} , \qq \qq \qq \: s'+i p'= R (s+ip)L^{\dagger} , \no \\
r'_\mu &=& R r_\mu R^{\dagger} + i R \partial_\mu R^{\dagger} , \qq
l'_\mu = L l_\mu L^{\dagger} + i L \partial_\mu L^{\dagger} , \no \\
\theta' &=& \theta + i \ln \det R - i \ln \det L ,
\end{eqnarray}
with $r_\mu = v_\mu + a_\mu, l_\mu = v_\mu - a_\mu$ and 
$R \in U(3)_R$, $L \in U(3)_L$, but the Lagrangian remains invariant. 
The phase of the determinant
det$U(x)=e^{i\psi}$ transforms  under axial $U(1)$ as
$\psi '= \psi  + 2 N_f \alpha$ so that
the combination $\psi + \theta$ remains invariant.
It is more convenient to replace the variable $\theta$ 
in the effective Lagrangian by this invariant
combination $\bar{\psi} = \psi + \theta$, ${\cal L}_{\mbox{\footnotesize 
eff}}= {\cal
L}_{\mbox{\footnotesize eff}}(U,\partial U,\ldots, v, a, s, p, \bar{\psi})$.
One can now construct the effective Lagrangian in these
fields that respects the symmetries of the underlying theory.
In particular, the Lagrangian is invariant under $U(3)_R \times U(3)_L$
rotations of $U$ and the external fields at a fixed value of the last argument.
The most general Lagrangian up to and including terms with two derivatives and
one factor of the quark mass matrix reads
\begin{eqnarray}  \label{mes}
{\cal L}^{(0+2)} &=& - V_0 + V_1 \langle \nabla_{\mu}U^{\dagger} \nabla^{\mu}U 
\rangle  + V_2 \langle U \chi^{\dagger} + U^{\dagger} \chi \rangle 
+ i V_3 \langle U \chi^{\dagger} - U^{\dagger} \chi \rangle \no \\
&&+ V_4 \langle U^{\dagger} \nabla_{\mu}U  \rangle 
\langle U^{\dagger} \nabla^{\mu}U 
\rangle + i V_5\langle U^{\dagger} \nabla_{\mu}U \rangle \nabla^{\mu} \theta
+ V_6 \nabla_{\mu} \theta \nabla^{\mu} \theta .
\end{eqnarray}
The expression $\langle \ldots \rangle$ denotes the trace in flavor space
and the quark mass matrix ${\cal M} = \mbox{diag}(m_u,m_d,m_s)$
enters in the combination
\begin{equation} \label{chi}
\chi = 2 B (s + i p) = 2 B {\cal M} 
\end{equation}
with $B = - \langle  0 | \bar{q} q | 0\rangle/ F^2$ the order
parameter of the spontaneous symmetry violation.
The covariant derivatives are defined by
\begin{eqnarray}
\nabla_{\mu} U  &=&  \partial_{\mu} U - i ( v_{\mu} + a_{\mu}) U
                     + i U ( v_{\mu} - a_{\mu})   \no \\
\nabla_{\mu} \theta  & = &  \partial_{\mu} \theta + 2 \langle a_{\mu} \rangle .
\end{eqnarray}
The coefficients $V_i$ are functions of the variable 
$\bar{\psi}$, $V_i(\bar{\psi})$,
and can be expanded in terms of this variable. At a given order of
derivatives of the meson fields $U$ and insertions of the quark mass matrix 
${\cal M}$ one obtains an infinite string of increasing powers of 
$\bar{\psi}$ with couplings which are not fixed by chiral symmetry.
Parity conservation implies that the $V_i$ are all even functions
of $\bar{\psi}$ except $V_3$, which is odd, and
$V_1(0) = V_2(0) = F^2/4$ gives the correct  normalizaton
for the quadratic terms of the Goldstone boson octet.
A transformation of the type $U \rightarrow e^{if(\bar{\psi})} U$
leaves the structure of the effective Lagrangian invariant,
but modifies the potentials $V_i$ \cite{KL}.
We will remove the term
$i \langle  U^{\dagger} \nabla_{\mu}U \rangle \nabla^{\mu} \theta$ 
by choosing $f(\bar{\psi})$ accordingly. 
Alternatively, one could simplify the potential $V_0(\bar{\psi})$ so
that it reads $V_0(\bar{\psi}) = v_0^{(0)} + v_0^{(2)} \bar{\psi}^2$. This is
achieved by replacing $\bar{\psi} \rightarrow \bar{\psi} e^{g(\bar{\psi})}$
while keeping $\theta$ fixed. The function
$g$ can be chosen in such a way that it cancels the terms with four and more
powers of $\bar{\psi}$ in the expansion of $V_0(\bar{\psi})$. 
On the other hand, the transformations for $U$ and $\bar{\psi}$
are related to each
other via $\psi = -i \ln \det U$, so that one cannot eliminate 
$V_5$ and simplify $V_0$ simultaneously. We prefer working with the Lagrangian
in which the potential $V_5$ has been transformed away and keep $V_0$.
Note also that $V_6$ does not contribute to the processes we are considering
here and will be neglected.

At fourth chiral order many more terms contribute \cite{H-S}, and we will only
present the relevant ones for our present investigation
\begin{equation}  
\begin{split}
        {\cal L}^{(4)} = & 
                \beta_4 \langle \nabla_\mu U^\dagger \nabla^\mu U\rangle \langle U^\dagger \chi + \chi^\dagger U\rangle 
                + \beta_5 \langle \nabla_\mu U^\dagger \nabla^\mu U (U^\dagger \chi + \chi^\dagger U)\rangle  \\
                & + \beta_6 \langle U^\dagger \chi + \chi^\dagger U\rangle ^2
                + \beta_7 \langle U^\dagger \chi - \chi^\dagger U\rangle ^2 \\
                & + \beta_8 \langle U^\dagger \chi U^\dagger \chi + \chi^\dagger U \chi^\dagger U\rangle 
                + \beta_{12} \langle \chi^\dagger \chi\rangle  \\
                & + \beta_{17} \langle U^\dagger \nabla_\mu U\rangle 
                                \langle U^\dagger \nabla^\mu U\rangle \langle U^\dagger \chi + \chi^\dagger U\rangle  \\
                & + \beta_{18} \langle U^\dagger \nabla_\mu U\rangle 
                                \langle \nabla^\mu U^\dagger \chi - \chi^\dagger \nabla^\mu U\rangle  \\
                & + i \beta_{25} \langle U^\dagger \chi U^\dagger \chi - \chi^\dagger U \chi^\dagger U\rangle  
                + i \beta_{26} ( \langle U^\dagger \chi\rangle ^2 - \langle \chi^\dagger U\rangle ^2 ) \\
                & - \beta_{52} \partial_\mu \nabla^\mu \theta \langle U^\dagger \chi + \chi^\dagger U\rangle 
\end{split}
\end{equation}   
The operators $O_{46}, O_{47}$ and $O_{53}$ from \cite{H-S} are not shown here,
since they can be removed via the transformation 
$U \rightarrow U \exp [ f_1 (\bar{\psi}) U^{\dagger} \nabla^\mu U 
\nabla_\mu \theta 
+ i f_2 (\bar{\psi}) \partial_\mu \nabla^\mu \theta 
+ f_3 (\bar{\psi}) \langle  U^{\dagger} \nabla_{\mu}U
\rangle \nabla^{\mu} \theta ] $ \cite{KL}.
We make use of this possibility and neglect these terms.
Furthermore, the vacuum angle has served its purpose in constructing the
effective Lagrangian and will be omitted for the processes under
consideration. However, one must keep the singlet component of
$a_\mu$, $\langle a_{\mu} \rangle $, in the covariant derivative
of $\nabla_{\mu} \theta  =  \partial_{\mu} 
\theta + 2 \langle a_{\mu} \rangle$.

\section{Identifying the singlet field}
In the last section we have mentioned that the singlet field $\eta_0$ is
described by the phase $\det U = e^{i \psi}$.
It remains to be seen, however, which choice of $\eta_0$ is sensible and how
this choice is related to $\psi$.
To this end, consider the dependence of the LECs in the effective 
Lagrangian on the
renormalization scale of QCD. This has been described in detail in
\cite{KL}. We will therefore just repeat the basic formulae and restrict
ourselves to the renormalization of the singlet axial current.
The renormalization of the scalar and pseudoscalar operators as well as the one
for the topological charge density $\omega =  ( g^2 / 16 \pi^2)
\, \mbox{tr}_c ( G_{\mu \nu} \tilde{G}^{\mu \nu} )$
is not relevant for our purposes and will be neglected.
The matrix elements of the singlet axial current $A_\mu^0 = \frac{1}{2}
\bar{q} \gamma_\mu \gamma_5 q$ depend on the renormalization scale of
QCD since this operator carries anomalous dimension \cite{ano}.
This operator receives multiplicative renormalization and, therefore, the decay
constants associated with the singlet quark current depend on the scale
\begin{equation}
A_\mu^{0} \rightarrow  Z_A A_\mu^0 , \qq \qq  F_P^{0}  \rightarrow 
Z_A F_P^0, \qq P = \eta, \eta'
\end{equation}
where the decay constants are given by
\begin{equation}
\langle 0| A_\mu^0|P \rangle = i p_\mu F_P^0 .
\end{equation}
The renormalization factor $Z_A$ depends on the running scale of QCD
\begin{equation}
\mu_{\scriptscriptstyle QCD} \frac{d Z_A}{d \mu_{\scriptscriptstyle QCD}} = 
\gamma_A Z_A, \qq
\gamma_A = - \frac{6 N_f (N_c^2 -1)}{N_c} \Big( \frac{g}{4 \pi} \Big)^4
+ {\cal O}(g^6)
\end{equation}
with $g$ being the QCD coupling constant.
Note that we work in the isospin limit $\hat{m} = m_u = m_d$, in which
$F_{\pi^0}^0$ vanishes.
It is convenient to compensate the scale dependence of the singlet axial
current by treating the corresponding external source as scale dependent field,
so that the effective action of QCD given by the Lagrangian in Eq. (\ref{lag})
becomes scale independent.
For $\theta = 0$ the effective action remains the same if we replace 
the singlet component of
the axial field by
\begin{equation}
\langle a_\mu \rangle \rightarrow   Z_A^{-1} \langle a_\mu \rangle 
\end{equation}
while the octet component $\hat{a}_\mu  = a_\mu - \langle a_\mu \rangle/3$ is
unaffected by a change in the renormalization scale.
Thus, if the vacuum angle $\theta$ is turned off, the situation for
$\langle a_\mu \rangle$ is analogous to the case of the scalar and pseudoscalar
external currents in standard chiral perturbation theory, where one introduces
the multiplicative constant $B$, cf. Eq. (\ref{chi}).
However, for a finite vacuum angle $\theta$, $\langle a_\mu \rangle$ is 
subject to an inhomogeneous renormalization \cite{KL}.
This complication does not arise here since we work with a vanishing vacuum
angle in the present investigation.
The scale invariance of the Lagrangian 
translates into the effective theory as follows.
For $\theta = 0$ the singlet component of the external axial field appears due
to axial $U(1)$ invariance only in the combination
$\nabla_\mu  \psi = \partial_\mu \psi - 2 \langle a_\mu \rangle$
which acquires multiplicative renormalization
\begin{equation}
\nabla_\mu  \psi  \rightarrow Z_A^{-1} \nabla_\mu  \psi , \qq
\mbox{i.e.} \q \psi  \rightarrow Z_A^{-1} \psi  .
\end{equation}
For the effective Lagrangian to remain invariant under a change of the
renormalization scale in QCD, the potentials $V_i$ and
$\beta_i$ must transform accordingly. 
We will illustrate this method in this section by
restricting ourselves to the Lagrangian ${\cal L}^{(0+2)}$ from Eq. 
(\ref{mes}).
The pertinent formulae for ${\cal L}^{(4)}$ can be found in App. A.

Decomposing the matrix valued field $U$ into its unimodular part and its phase
\begin{equation}
U = e^{\frac{i}{3}\psi} \hat{U}
\end{equation}
and the axial-vector field $a_\mu$ into octet and singlet components
\begin{equation}
a_\mu = \hat{a}_\mu + \frac{1}{3} \langle a_\mu \rangle , 
\end{equation}
the Lagrangian ${\cal L}^{(0+2)}$ can be rewritten as ($V_5 = V_6 = 0$),
\begin{eqnarray}  \label{mes1}
{\cal L}^{(0+2)} &=& - V_0 + V_1 \langle \hat{\nabla}_{\mu} 
\hat{U}^{\dagger} \hat{\nabla}^{\mu}\hat{U} \rangle  + 
[V_2 + i V_3 ] e^{\frac{i}{3}\psi} \langle \hat{U} 
\chi^{\dagger}\rangle \no \\
&&+ [V_2 - i V_3 ] e^{-\frac{i}{3}\psi} \langle 
\hat{U}^{\dagger} \chi \rangle 
+[ \frac{1}{3} V_1 - V_4 ] \nabla_{\mu} \psi \nabla^{\mu} \psi  .
\end{eqnarray}
The covariant derivative of $\hat{U}$ is given by
\begin{equation}
\hat{\nabla}_{\mu} \hat{U}  =  \partial_{\mu} \hat{U} 
     - i ( v_{\mu} + \hat{a}_{\mu}) \hat{U} 
   + i \hat{U} ( v_{\mu} - \hat{a}_{\mu}).
\end{equation}
For the Lagrangian
to remain invariant, the potentials $V_i$ must transform as
\begin{eqnarray}
V_0 (x)   & \rightarrow &   V_0 (Z_A x) \no \\
V_1 (x)   & \rightarrow &   V_1 (Z_A x) \no \\
(V_2 + i V_3) (x)   & \rightarrow &  (V_2 + i V_3)  (Z_A x) \,
e^{\frac{i}{3} (Z_A -1) x}  \no \\
V_4 (x)   & \rightarrow &  Z_A^2 V_4(Z_A x) +\frac{1}{3} (1-Z_A^2) V_1(Z_A x).
\end{eqnarray}
These transformation properties of the potentials have consequences 
for the choice of
the singlet field. Consider the free kinetic term for $\psi$ in
Eq. (\ref{mes1})
\begin{equation}
[ \frac{1}{3} V_1(0) - V_4(0) ]  \partial_{\mu} \psi \partial^{\mu} \psi .
\end{equation}
Both the coefficient $V_1(0)/3 -V_4(0)$ and $\psi$ are scale dependent
quantities with
\begin{equation}
 \frac{1}{3} V_1(0) - V_4(0) = \frac{F^2}{12}  - V_4(0)
\rightarrow  Z_A^2 \Big( \frac{F^2}{12}  - V_4(0) \Big).
\end{equation}
Note that $F$, the pion decay constant in the chiral limit, does not
depend on the running scale of QCD.
We prefer to work with a scale independent singlet field $\eta_0$
which has the same kinetic term as the octet fields. 
This is achieved by replacing $\psi$ by $\eta_0$ in the effective
Lagrangian with
\begin{equation}
\eta_0 = \sqrt{\lambda} \, \psi \equiv  
    \sqrt{\frac{F^2}{6}  - 2 V_4(0) }  \, \psi .
\end{equation}
The potentials are then rescaled according to
\begin{eqnarray}
\bar{V}_0 (x)  & = &   V_0 \Big ( \frac{F}{\sqrt{\lambda}} x \Big)   \no \\
\bar{V}_1 (x)   & = &   V_1 \Big ( \frac{F}{\sqrt{\lambda}} x \Big) \no \\
(\bar{V}_2 + i \bar{V}_3) (x)   & = &  (V_2 + i V_3) 
\Big ( \frac{F}{\sqrt{\lambda}} x \Big)  \,
e^{ \frac{i}{3} (F -\sqrt{6 \lambda} ) x/\sqrt{\lambda}} \no \\
(\frac{1}{3} \bar{V}_1 - \bar{V}_4)(x)  & = &  \frac{F^2}{6 \lambda}
( \frac{1}{3} V_1 - V_4) \Big ( \frac{F}{\sqrt{\lambda}} x  \Big).
\end{eqnarray}
The $\bar{V}_i$ are functions of the singlet field $\eta_0$ , 
$\bar{V}_i(\eta_0/F)$, and do not depend on the renormalization scale
$\mu_{\scriptscriptstyle QCD}$.
On the other hand, $\nabla_\mu \psi$ transforms into
\begin{equation}
\nabla_\mu \psi \rightarrow \frac{1}{\sqrt{\lambda}} \nabla_\mu \eta_0
= \frac{1}{\sqrt{\lambda}} \Big( \partial_\mu \eta_0 - 2 \sqrt{\lambda} 
\langle a_\mu \rangle \Big)
\end{equation}
so that
\begin{equation}
\Big( \frac{1}{3} V_1 - V_4\Big) (\psi) \nabla_{\mu} \psi \nabla^{\mu} \psi
\q \rightarrow \q  
\frac{6}{F^2} \Big(\frac{1}{3} \bar{V}_1 - \bar{V}_4\Big)(\frac{\eta_0}{F}) 
\nabla_{\mu} \eta_0  \nabla^{\mu} \eta_0 .
\end{equation}
Note that the kinetic term for $\eta_0$ which is the first term in the
expansion of $\bar{V}_1/3 - \bar{V}_4$ is normalized in such a way that 
$\bar{V}_1(0)/3 - \bar{V}_4(0)= F^2/12$, i.e. $\bar{V}_4(0)=0$.
In the effective theory the scale dependence of the axial current manifests
itself in the prefactor $\sqrt{\lambda}$ of $\langle a_\mu \rangle$.
We also would like to point out that the quantity
$\lambda = F^2/6 - 2 V_4(0)$ is indeed a positive number.
This can be seen as follows.
The pieces of the Lagrangian in Eq. (\ref{mes1}) quadratic in the field 
$\psi$ can be written in the chiral
limit as
\begin{equation} \label{quad}
\frac{1}{2} \Big[ \frac{F^2}{6} - 2 V_4(0) \Big] \partial_\mu \psi 
\partial^\mu \psi - v_0^{(2)} \psi^2 
\end{equation}
where the first and second term constitute the kinetic energy and the mass
term, respectively. 
At lowest order in $1/N_c$ the equality $2 v_0^{(2)} = 
\tau_{\scriptscriptstyle GD}$ holds, with $\tau_{\scriptscriptstyle GD}$ being
the topological susceptibility of gluodynamics \cite{KL}.
The domain of validity for standard $SU(3)$ chiral perturbation theory
is restricted by the condition \cite{leu2}
\begin{equation}
m_s | \langle 0| \bar{u} u| 0 \rangle| \ll 9 \tau_{\scriptscriptstyle GD} .
\end{equation}
Assuming small $1/N_c$ corrections, $v_0^{(2)}$ is thus a positive number.
For the free effective Lagrangian in Eq. (\ref{quad}) to make sense,
we obtain the constraint
\begin{equation}
\frac{F^2}{6} - 2 V_4(0) > 0 .
\end{equation}
Otherwise, the corresponding $\psi$ propagator does not develop a pole and no
singlet particle occurs.

We will now apply the Lagrangian with the octet $\phi$ and $\eta_0$ to
calculate both the meson masses at lowest order and $\eta_0$-$\eta_8$ mixing,
which yields the physical states $\eta$ and $\eta'$.
The potentials $\bar{V}_i$ are expanded in the singlet field $\eta_0$ 
\begin{eqnarray}
\bar{V}_i(\frac{\eta_0}{F}) &=& \bar{v}_i^{(0)} + \bar{v}_i^{(2)} 
\frac{\eta_0^2}{F^2} +
\bar{v}_i^{(4)} \frac{\eta_0^4}{F^4} + \ldots
\qq \mbox{for} \q i= 0,1,2,4 \no \\
\bar{V}_3(\frac{\eta_0}{F}) &=& \bar{v}_3^{(1)} \frac{\eta_0}{F} + 
\bar{v}_3^{(3)} \frac{\eta_0^3}{F^3}  
+ \ldots \q .
\end{eqnarray}
The expansion coefficients $\bar{v}_i^{(j)}$ are independent of the running
scale of QCD, whereas in general
the corresponding coefficients $v_i^{(j)}$ of the potentials $V_i$ are not.
One observes terms quadratic in the meson fields that contain the factor
$\eta_0 \eta_8$. Such terms arise from the explicitly chiral symmetry breaking
operators $(\bar{V}_2 + i \bar{V}_3) e^{i \sqrt{6} \eta_0 /(3 F)}
\langle \hat{U} \chi^\dagger \rangle + h.c.$
and read
\begin{equation} \label{mix1}
- \frac{8}{\sqrt{3} F^2} 
\Big[ \frac{F^2}{2 \sqrt{6}} + \bar{v}_3^{(1)} \Big] B (\hat{m}
  - m_s) \eta_0 \eta_8 \equiv - m_{08}^2 \eta_0 \eta_8 .
\end{equation}
The states 
$\eta_0$ and $\eta_8$ are not mass eigenstates and the mass matrix can
be diagonalized by introducing the eigenstates $\eta$ and $\eta'$
\begin{eqnarray} \label{mix2}
| \eta \rangle & = & \cos \vartheta \, | \eta_8 \rangle -
                     \sin \vartheta \, | \eta_0 \rangle  \no \\
| \eta'\rangle & = & \sin \vartheta \, | \eta_8 \rangle +
                     \cos \vartheta \, | \eta_0 \rangle . 
\end{eqnarray}
Eqs. (\ref{mix1}) and (\ref{mix2}) can be used to extract numerical values for
$\bar{v}_3^{(1)}$ as a function of the mixing angle $\vartheta$. 
To this end, we use the relation $B (\hat{m}
  - m_s) = m_\pi^2 - m_K^2$ which is valid at lowest chiral order,
cf. Eq. (\ref{masses}).
For $\vartheta = 0^\circ$ one obtains then $\bar{v}_3^{(1)} = -1.77 \times
 10^{-3}$ GeV$^2$, whereas $\vartheta = -20^\circ$ \cite{pdg} leads to 
$\bar{v}_3^{(1)} = -0.76 \times 10^{-3}$ GeV$^2$.
The relation between the $\eta$-$\eta'$ mixing angle $\vartheta$ and 
$\bar{v}_3^{(1)}$ in terms of the meson masses
will change, of course, at higher chiral orders. We will
therefore not treat $\bar{v}_3^{(1)}$ as a function of $\vartheta$ and leave
its numerical value undetermined in the present investigation.

Up to second chiral order the masses for the pseudoscalar mesons
read (with a mixing angle $\vartheta = 0^\circ$)
\begin{eqnarray}  \label{masses}
{\stackrel{\circ}{m}}_\pi^2 &=& 2 B \hat{m} \no \\
{\stackrel{\circ}{m}}_K^2 &=& B [ \hat{m} + m_s] \no \\
{\stackrel{\circ}{m}}_\eta^2 &=& \frac{2}{3} B [\hat{m} + 2 m_s] \no \\
{\stackrel{\circ}{m}}_{\eta'}^2 &=& 
\frac{2}{F^2} \bar{v}_0^{(2)} - \frac{8}{F^2} B 
[ 2 \hat{m} + m_s] \Big( \bar{v}_2^{(2)}
- \frac{F^2}{12} - \sqrt{\frac{2}{3}} \bar{v}_3^{(1)} \Big)
\end{eqnarray}
with ${\stackrel{\circ}{m}}_P$ denoting the leading terms in the expansion of
the masses. Note the $\bar{v}_0^{(2)}$ term in the expression for
${\stackrel{\circ}{m}}_{\eta'}$ which does not vanish in the chiral limit, 
i.e. the $\eta'$ is not a Goldstone boson.
The mixing angle $\vartheta$ is given by
\begin{equation}
\tan 2 \vartheta = \frac{2 m_{08}^2}{{\stackrel{\circ}{m}}_{\eta'}^2 - 
{\stackrel{\circ}{m}}_\eta^2} .
\end{equation}
It can be expanded in powers of 1/${\stackrel{\circ}{m}}_{\eta'}^2$ 
so that one arrives at
\begin{equation} \label{the}
\tan 2 \vartheta =  \frac{2 m_{08}^2}{{\stackrel{\circ}{m}}_{\eta'}^2} 
+ {\cal O}(p^4) .
\end{equation}
In this framework the generalized Gell-Mann-Okubo mass relation for the
pseudoscalar mesons reads
\begin{equation} \label{gmo}
\sin^2 \vartheta \, {\stackrel{\circ}{m}}_{\eta'}^2 + \cos^2 \vartheta \, 
{\stackrel{\circ}{m}}_\eta^2 =
\frac{1}{3} ( 4 {\stackrel{\circ}{m}}_K^2 - {\stackrel{\circ}{m}}_\pi^2)
\end{equation}
which reduces to the conventional formula if one uses from 
Eq. (\ref{the}) $\, \vartheta \sim {\cal O}(p^2)$, i.e. $\cos \vartheta \sim 1$ and 
$\sin \vartheta \sim 0$.
For the physical values of the meson masses the above relation (\ref{gmo}) 
yields $|\vartheta| \approx 10^\circ$.
We have now established the theory at lowest order 
and can proceed by calculating the loop contributions
in the following section.

\section{Masses and decay constants}
We have set up the effective Lagrangian for the calculation of the masses and 
decay constants of the pseudoscalar meson nonet up to fourth chiral order,
i.e. one-loop order. The next step is to calculate the contributions from
chiral loops. Employing dimensional regularization for the loop integrals
amounts to a chiral expansion in $m_P^2/\Lambda^2_\chi$, $P= \pi, K, \eta,
\eta'$. While the expansion parameters are small for the Goldstone boson octet,
it is close to unity for the $\eta'$ with $m_{\eta'}/\Lambda_\chi 
\sim 0.8$, and the convergence of the series is doubtful.
The mass of the $\eta'$ is a quantity of zeroth chiral order,
$m_{\eta'} \sim {\cal O}(p^0)$, and this will spoil the chiral counting scheme:
higher loop graphs will also contribute to lower chiral orders.
The situation is thus similar to baryon chiral perturbation theory 
where the nucleon mass sets a scale of similar size.

In \cite{BL} a new regularization method has been introduced, the so-called
infrared regularization, which preserves the chiral counting scheme in the
presence of massive fields, while keeping Lorentz and chiral invariance
explicit at all stages. This method has so far only been employed in
baryon chiral perturbation theory, see e.g.  \cite{BL, ET2, MK}, 
but is applicable for any massive particle.
In this section, we present the calculation of the masses and decay constants
of the meson nonet up to one-loop order in infrared regularization. 
We will compare our results with the expressions obtained in dimensional
regularization. 
At one-loop order only the tadpole graph contributes. The fundamental loop
integral is given by
\begin{equation}
I = \int \frac{d^d k}{(2 \pi)^d} \frac{i}{k^2 - m_P^2 + i \epsilon}
\end{equation}
with $m_P$ being the mass of the meson inside the loop.
We use the physical mass which is consistent to the order we are working.
According to \cite{BL} the integral can be decomposed into a regular and a
singular part, $I = S + R$, where both the pieces $S$ and $R$ preserve the
symmetries of the Lagrangian. The regular part is a polynomial in the quark
masses and can thus be absorbed by a suitable redefiniton of the LECs.
We will drop the regular part of this integral and keep the singular components
which are given by
\begin{eqnarray}
S_\phi &=& m_\phi^2 \Big[ 2 L + \frac{1}{16 \pi ^2} \ln \frac{m_\phi^2}{\mu^2}
\Big] \qq \phi = \pi, K, \eta \no \\
S_{\eta'} &=& 0
\end{eqnarray}
with
\begin{equation}
L = \frac{\mu^{d-4}}{16 \pi^2} \Big\{ \frac{1}{d-4} - \frac{1}{2} 
   [ \ln 4 \pi + 1 - \gamma_E] \Big\}.
\end{equation}
and $\gamma_E = 0.5772...$ is the Euler-Mascheroni constant.
The quantity $\mu$ is the scale introduced in the regularization of the
integral. The divergent pieces of the chiral invariant singular parts
constitute polynomials in the quark masses and can be cancelled by an
appropriate renormalization of the coupling constants.
Our main concern in this introductory paper is the applicability of $U(3)$
chiral perturbation theory to phenomenology, in particular the convergence of
the chiral series. The more technical issue of the renormalization prescription
for the LECs $\beta_i$ of ${\cal L}^{(4)}$ will be neglected.
(The complete renormalization of the one-loop functional in dimensional
regularization has been given in \cite{H-S}.)
We will therefore assume that both the regular part and the divergences from
the singular part have been absorbed by a redefinition of the LECs and will use
the same notation for the renormalized coupling constants. In our calculation
this amounts to keeping only the chiral logarithms of the loops with the
Goldstone bosons,
\begin{equation}
S_\phi \rightarrow \frac{1}{16 \pi ^2} m_\phi^2 \ln \frac{m_\phi^2}{\mu^2}
\qq \phi = \pi, K, \eta .
\end{equation}

The calculation of the masses for the meson nonet up to fourth chiral order
including one-loop corrections from ${\cal L}^{(0+2)}$ then yields
\begin{equation}  \label{mass}
m_P^2 = {\stackrel{\circ}{m}}_P^2 + \frac{1}{F^2} C^{ab}_P (\mu) m_a^2 m_b^2 +
\frac{1}{16 \pi ^2 F^2} D^{ab}_P   m_a^2 m_b^2 \ln \frac{m_b^2}{\mu^2}
+ \Delta_P (\mu)
\end{equation}
with $P = \pi, K, \eta , \eta'$.
The counterterms contained in $ C^{ab}_P$ cancel the $\mu$-dependence of
the chiral logarithms and $\Delta_P  $ 
includes both the corrections to ${\stackrel{\circ}{m}}_{\eta, \eta'}^2$ 
due to $\eta$-$\eta'$ mixing
and the tadpole contribution of the $\bar{v}_0^{(4)}$ vertex.
The explicit expressions for $ C^{ab}_P $, $D^{ab}_P$ and
$\Delta_P $ can be found in
App. B.

The generalization to the corresponding expressions in dimensional
regularization is straightforward. It is exclusively the tadpole contribution
of the $\eta'$ which adds to the nonanalytic pieces, so that for the masses
$m_P^{2 (dim)} $ in dimensional regularization we can write
\begin{equation}  \label{mass2}
m_P^{2 (dim)} = m_P^2 + 
\frac{1}{16 \pi ^2 F^2} D^{a\eta'}_P   m_a^2 m_{\eta'}^2 \ln 
\frac{m_{\eta'}^2}{\mu^2} + \Delta_P^{(dim)} (\mu)
\end{equation}
with $m_P^2$ given in Eq. (\ref{mass}).
The $D^{a\eta'}_P$ and $\Delta_P^{(dim)}$ are given in App. B.
Of course, the difference between both schemes vanishes for $\mu = m_{\eta'}$
which is a peculiarity of the one-loop calculation of the masses since only the
tadpole contributes. The degeneracy of both regularization schemes for
$\mu = m_{\eta'}$ disappears in a two-loop calculation or for other processes
which yield different nonanalytic contributions. An investigation of this
is beyond the scope of the present work.
Here, we evaluate the difference of both regularization schemes by varying the
scale $\mu$ between the $\rho$ mass, $m_{\rho} = 770$ MeV, and $\Lambda_\chi
\sim 1.2$ GeV.
Furthermore, the values of most of the couplings are not known and have to be
determined in principle from experiment. In order to obtain numerical results
we set all the couplings equal to zero except those from standard chiral
perturbation theory, i.e. $\bar{v}_1^{(0)} = \bar{v}_2^{(0)} = F^2/4$
at second chiral order and the kinetic term of the $\eta_0$ has been normalized
as explained above by using $\bar{v}_1^{(0)}/3 - \bar{v}_4^{(0)} = F^2/12$.
At fourth chiral order
we use the central phenomenological values for the renormalized
$\beta_i^{(0)}(m_\rho)$ 
as given in \cite{BEG}. These are in units of 10$^{-3}$:
$\bar{\beta}_4^{(0)} = -0.3, \bar{\beta}_5^{(0)} = 1.4, \bar{\beta}_6^{(0)} 
=  -0.2,  \bar{\beta}_7^{(0)} = -0.4$ and 
$\bar{\beta}_8^{(0)} = 0.9$.
The couplings $\bar{\beta}_{17}^{(0)}$ and $\bar{\beta}_{18}^{(0)}$ are 
OZI-violating corrections, while 
$\bar{\beta}_{25,26,52}^{(1)}$ are parity violating operators.
Resonance exchange calculations as performed in \cite{EGPR} yield vanishing
values for $\bar{\beta}_{17}^{(0)} \ldots \bar{\beta}_{52}^{(1)}$ 
since the resonance couplings used within
this approach obey both the OZI-rule and are parity conserving. We will
therefore set the values of these counterterms equal to zero.
Table 1 shows the dependence of the next-to-leading order mass contributions 
on the scale $\mu$ 
for the values $\mu = m_{\rho} = 770$ MeV, $\mu = m_{\eta'} = 958$ MeV and  $\mu = \Lambda_\chi= 1.2$ GeV,
where we have used an 
$\eta$-$\eta'$ mixing angle of both $\vartheta = 0^\circ$ and
$\vartheta = -20^\circ$ \cite{pdg}.
In the calculation of the higher chiral orders we have kept for convenience
the $\bar{\beta}_i$ at the scale $\mu = m_{\rho}$, $\bar{\beta}_i(m_{\rho})$.
Then the dependence of the meson masses on the renormalization scales stems
only from the chiral logarithms and is rather weak.
This in turn implies that the scale dependence of the $\bar{\beta}_i$
which should compensate this effect is also rather weak.
We also present the numerical dependence of the masses on the unknown couplings
$\bar{v}_i^{(j)}$ 
not known from standard chiral perturbation theory in Table 2.
This table should be read as follows: e.g. the fourth order contribution to the
$\eta'$ mass is given as (in units of GeV$^2$ and $\vartheta = -20^\circ, \mu =
m_\rho=0.77$ GeV)
\begin{eqnarray}
m^2_{\eta'} - \overcirc{m}^2_{\eta'} &=&  -0.05601 -21.47 \bar{v}_0^{(4)}
+ 90.15 \bar{v}_1^{(2)}
- 58.41 \bar{v}_2^{(2)} + 21.80 \bar{v}_2^{(4)} \nonumber \\
&& + 148.22 \bar{v}_3^{(1)}
+ 8.73 \bar{v}_3^{(3)}  - 26.13 \bar{v}_4^{(2)} 
\end{eqnarray}
in infrared regularization and
\begin{eqnarray}
m^2_{\eta'} - \overcirc{m}^2_{\eta'} &=&  -0.11904 + 296.14 \bar{v}_0^{(4)}
- 129.92 \bar{v}_1^{(2)}
+ 173.76 \bar{v}_2^{(2)} - 300.69 \bar{v}_2^{(4)} \nonumber \\
&& + 53.31 \bar{v}_3^{(1)}
+ 152.20 \bar{v}_3^{(3)}  + 556.86 \bar{v}_4^{(2)} 
\end{eqnarray}
in dimensional regularization.
The first number denotes the fourth order contribution from the couplings of
standard chiral perturbation theory as given in Table 1.
From these results it becomes obvious that the nonanalytic pieces of the
$\eta'$ loops proportional to the unknown couplings $\bar{v}_i^{(j)}$
are numerically much more significant in dimensional regularization 
than in infrared regularization.
Using dimensional regularization will eventually lead to a breakdown of the
chiral expansion. Since in this scheme higher loops with more
$\eta'$-propagators   correspond to higher powers in $m_{\eta'}^2$, we cannot
expect the chiral series to converge. This can be prevented by using infrared
regularization similar to the situation with nucleons in baryon chiral
perturbation theory.

We now turn to the calculation of the pseudoscalar decay constants. They are
defined by
\begin{equation}
\langle 0| A_\mu^a|P \rangle = i p_\mu F_P^a  \qq P = \pi, K, \eta, \eta'
\end{equation}
with $A_\mu^a = \bar{q} \gamma_\mu \gamma_5 \frac{1}{2} \lambda_a q$
and $\langle \lambda^a \lambda^b \rangle = 2 \delta^{ab}$.
One introduces the parametrization
\begin{eqnarray}
F^8_\eta &=& \cos \vartheta_8 F_8, \qq   F^0_\eta = -\sin \vartheta_0 F_0, \no \\
F^8_{\eta'} &=& \sin \vartheta_8 F_8, \qq   F^0_{\eta'} = \cos \vartheta_0 F_0 ,
\end{eqnarray}
which upon inversion leads to
\begin{equation}
(F_8)^2 = (F^8_\eta)^2 + (F^8_{\eta'})^2 , \qq 
(F_0)^2 = (F^0_\eta)^2 + (F^0_{\eta'})^2 .
\end{equation}
The results for the decay constants can be written in the form
\begin{equation}  \label{dec}
F_\phi= F \Big( 1 + \frac{1}{F^2} G^{a}_\phi (\mu) m_a^2  +
\frac{1}{16 \pi ^2 F^2} H^{a}_\phi   m_a^2 \ln \frac{m_a^2}{\mu^2} \Big)
\end{equation}
with $\phi = \pi, K, 8$ and the counterterms contained in $G^{a}_\phi$ have
absorbed the divergences from the loops.
The coefficients $G$ and $H$ can be found in App. C together with the pertinent
$Z$-factors.
Due to the appearance of many unknown LECs we will proceed similar to the case
of the masses by separating the known chiral logarithms and
coupling constants as given from standard chiral perturbation theory
from the undetermined couplings $\bar{v}_i^{(j)}$.
Table 3 shows in analogy to Table 2 the dependence of the pseudoscalar decay
constants on the unknown parameters $\bar{v}_i^{(j)}$ where we have chosen $\mu
= m_\rho$ and $\vartheta = -20^\circ$. 
For $F_8$, e.g., we obtain (in units  of GeV)
\begin{equation}
F_8 = F + 0.0444 - 0.3769 \bar{v}_1^{(2)}   - 0.2336 \bar{v}_4^{(2)} 
\end{equation}
in infrared regularization whereas the pertinent result in dimensional
regularization reads
\begin{equation}
F_8 = F + 0.0444 + 5.1983 \bar{v}_1^{(2)} + 3.2217 \bar{v}_4^{(2)} 
\end{equation}
with the first number being the contribution from the known LECs of standard
chiral perturbation theory.
Again, the $\eta'$ loops lead to significant
contributions in dimensional regularization.

The numerical value for $F_0$ cannot be determined from experiment
since it depends on the running scale of QCD.
The result for $F_0$ may be written as
\begin{equation}  \label{dec2}
F_0 = \sqrt{6 \lambda} \Big( \, 1
+ \frac{1}{F^2} G^{a}_0 (\mu) m_a^2  +
\frac{1}{16 \pi ^2 F^2} H^{a}_0   m_a^2 \ln \frac{m_a^2}{\mu^2} \Big)
\equiv \frac{\sqrt{6  \lambda } }{F}  \bar{F}_0
\end{equation}
where $\bar{F}_0$ is scale invariant.
The pertinent coefficients are given in App. C
and the higher order contributions for $\bar{F}_0$ are shown in Table 3.
They read in units of GeV
\begin{equation}
\bar{F}_0 = F - 0.0060 - 0.3769 \bar{v}_1^{(2)}   + 1.3643 \bar{v}_4^{(2)} 
\end{equation}
in infrared regularization and
\begin{equation}
\bar{F}_0 = F - 0.0060 + 5.1983 \bar{v}_1^{(2)}   -18.8266 \bar{v}_4^{(2)} .
\end{equation}
in dimensional regularization. The first number includes the couplings known
from standard chiral perturbation theory with all the remaining parameters set
equal to zero.

Finally, we would like to comment on the numerical values of the angles
$\vartheta_0$ and $\vartheta_8$. Their exact values up to one-loop order cannot be
extracted since some of the couplings are unknown. We will therefore proceed in
analogy to the masses and decay constants by using the phenomenological values
for the LECs of standard chiral perturbation theory and by neglecting the
remaining ones.
Using the identities
\begin{equation}
\tan \vartheta_8 = \frac{F_{\eta'}^8}{F_{\eta}^8} , \qq
\tan \vartheta_0 = - \frac{F_{\eta}^0}{F_{\eta'}^0}
\end{equation}
we obtain $\vartheta_0 = -4.42^\circ $ and $\vartheta_8 = -30.01^\circ$
for an $\eta$-$\eta'$ mixing angle of $-20^\circ $.
In order to obtain an estimate of the uncertainty that results from using
different values of the $\eta$-$\eta'$ mixing angle, we perform the same
calculation by employing $\vartheta =-13^\circ $ as given in \cite{BES}.
The pertinent angles $\vartheta_0$ and $\vartheta_8$ are then
$\vartheta_0 = 3.26^\circ $ and $\vartheta_8 = -24.06^\circ$.

\section{Summary}
In this investigation, we have presented an effective field theory which
describes the interactions of the Goldstone boson octet with the corresponding
singlet $\eta'$ without imposing 1/$N_c$ counting rules. The method has been
illustrated by calculating the masses and decay constants of the pseudoscalar
meson nonet up to one-loop order. The relevant effective Lagrangian up to
fourth chiral order has been given. It turns out -- as already discussed in
\cite{KL} -- that the LECs and the singlet field itself depend on the running
scale of QCD. Rescaling the singlet field however yields QCD scale invariant
coupling constants and the only scale dependent quantity of the effective
theory shows up as a prefactor of the singlet axial current. This is in
complete agreement with QCD since the axial vector current has anomalous
dimension and acquires multiplicative renormalization. 

Since the mass of the $\eta'$, $m_{\eta'}$, is close to the scale of chiral
symmetry breaking, $\Lambda_\chi$, 
dimensional regularization is not well suited for performing
the loop integration. It yields an expansion in $m_{\eta'}/\Lambda_\chi$, 
thus causing the
breakdown of the chiral expansion. This can be prevented by using so-called
infrared regularization which suppresses the $\eta'$ contribution to the
amplitudes \cite{BL}. 
In the present work,
the nonanalytic pieces of the one-loop integrals have been compared between
both regularization schemes and it has been found that in dimensional
regularization the $\eta'$ tadpole leads to significant contributions,
suggesting that the expansion in $m_{\eta'}$ will not be as well-behaved as for
the Goldstone boson octet. In dimensional regularization one cannot expect
higher loops to be less significant.
A peculiarity of the calculation of the masses and decay constants is that the
$\eta'$ loop contributions vanish identically in infrared regularization, since
at one-loop order only the tadpole contributes.
This will change if one goes to higher loop order or considers other processes
with different nonanalytic contributions.
Nevertheless, the convergence of the chiral series will be improved by using
infrared regularization as it is the case in baryon chiral perturbation
theory. In order to confirm our results other processes such as the hadronic
decay modes of the $\eta$ and $\eta'$ will be investigated within this
framework in future studies.

\section*{Acknowledgments}
We would like to thank N. Beisert for valuable discussions and a very careful
reading of the manuscript.

\newpage

\appendix 
\def\theequation{\Alph{section}.\arabic{equation}}
\setcounter{equation}{0}
\section{} \label{app:a}
In this Appendix, we present the dependence of the potentials $\beta_i$
on the running scale of QCD.
To this end, the Lagrangian ${\cal L}^{(4)}$ is rewritten in terms of
$\hat{U}, \hat{a}_\mu, \nabla_\mu \psi$ $ (\theta =0)$
\begin{equation}
\begin{split}
        {\cal L}^{(4)} = & 
  ( \beta_4 -i \beta_{22} ) e^{-\frac{i}{3} \psi} 
  \langle \hat{\nabla}_\mu \hat{U}^\dagger \hat{\nabla}^\mu \hat{U}\rangle 
                    \langle \hat{U}^\dagger \chi\rangle  + h.c.\\
      &+( \beta_5 -i \beta_{21} ) e^{-\frac{i}{3} \psi} 
         \langle \hat{\nabla}_\mu \hat{U}^\dagger \hat{\nabla}^\mu \hat{U}
                \, \hat{U}^\dagger \chi\rangle  + h.c.\\
                                &+\frac{i}{3} ( 2\beta_5 + 3 \beta_{18} -2 i \beta_{21} +3 i \beta_{23} )
                                                e^{-\frac{i}{3} \psi} \nabla_\mu \psi
                                                \langle \hat{\nabla}^\mu \hat{U}^\dagger \chi\rangle  + h.c.\\
                                &+\frac{1}{9} ( 3 \beta_4 + \beta_5 -9 \beta_{17} +3 \beta_{18} -i \beta_{21}
                                                        -3 i \beta_{22} +3 i \beta_{23} -9 i \beta_{24} )\\
                                    & \qquad \qquad \times e^{-\frac{i}{3}
                                    \psi} \nabla_\mu \psi \nabla^\mu \psi
                                                        \langle \hat{U}^\dagger \chi\rangle  + h.c.\\
                                &+( \beta_6 + \beta_7 +i \beta_{26} ) e^{-\frac{2 i}{3} \psi} 
                                                \langle \hat{U}^\dagger \chi\rangle ^2 + h.c.\\
                                &+2 ( \beta_6 - \beta_7 ) \langle \hat{U}^\dagger \chi\rangle 
                                                \langle \chi^\dagger \hat{U}\rangle 
                                 + \beta_{12} \langle  \chi^\dagger \chi \rangle  \\
                                &+(\beta_8 -i \beta_{25} ) e^{\frac{2 i}{3} \psi} 
                                                \langle \chi^\dagger \hat{U}
                                                \chi^\dagger \hat{U}\rangle  +
                                                h.c. \\
                                &-(\beta_{52} -i \beta_{53}) e^{-\frac{i}{3} \psi} 
                                                \partial_\mu \nabla^\mu \theta
                                                \langle \hat{U}^\dagger
                                                \chi\rangle  + h.c.
\end{split}
\end{equation}
where we have also added the operators
${\cal O}_i, i= 21,22,23,24,53$.
Although they do not contribute to the masses and decay constants, they are
needed to reveal the scale dependence of the potentials.
In order for the effective Lagrangian to remain invariant under a change of the
QCD scale, the potentials $\beta_i$ have to transform as
\begin{equation}
\begin{split}
        (\beta_4 - i \beta_{22})(x) \to & (\beta_4 - i \beta_{22})(Z_A x) e^{\frac{i}{3} (1-Z_A) x} \\
        (\beta_5 - i \beta_{21})(x) \to & (\beta_5 - i \beta_{21})(Z_A x) e^{\frac{i}{3} (1-Z_A) x} \\
        ( 2\beta_5 + 3 \beta_{18} -2 i \beta_{21} +3 i \beta_{23} )(x) \to &
                        ( 2\beta_5 + 3 \beta_{18} -2 i \beta_{21} +3 i \beta_{23} )(Z_A x)\\ 
                                        & \qquad \times e^{\frac{i}{3} (1-Z_A) x} Z_A \\
        ( 3 \beta_4 + \beta_5 -9 \beta_{17} +3 \beta_{18} -i \beta_{21} \qquad & \\
                        -3 i \beta_{22} +3 i \beta_{23} -9 i \beta_{24} )(x) \to& 
                                ( 3 \beta_4 + \beta_5 -9 \beta_{17} +3 \beta_{18} -i \beta_{21} \\
                                & \qquad -3 i \beta_{22} +3 i \beta_{23} -9 i \beta_{24} )(Z_A x) \\
                                        & \qquad \qquad \times e^{\frac{i}{3} (1-Z_A) x} Z^2_A \\
        (\beta_6 + \beta_7 + i \beta_{26})(x) \to & (\beta_6 + \beta_7 + i \beta_{26})(Z_A x) 
                                e^{\frac{2 i}{3} (1-Z_A) x} \\
        (\beta_6 - \beta_7)(x) \to & (\beta_6 - \beta_7)(Z_A x) \\
        \beta_{12} (x) \to & \beta_{12} (Z_A x) \\
        (\beta_8+ i \beta_{25})(x) \to & (\beta_8 + i \beta_{25})(Z_A x) 
                                e^{\frac{2 i}{3} (1-Z_A) x} \\
        (\beta_{52} -i \beta_{53})(x) \to & (\beta_{52} -i \beta_{53})(Z_A x) 
                                e^{\frac{i}{3} (1-Z_A) x} Z_A .
\end{split}
\end{equation}
Note that the term $\beta_{12} \langle \chi \chi^\dagger \rangle$ 
contains a contact term,$\beta_{12}(0) \langle \chi \chi^\dagger \rangle$,
which involves the renormalization of the corresponding counterterm in QCD, 
$ \beta \langle s s^\dagger \rangle$.
Such a term is consistent with the symmetries of QCD and is needed to render
the effective action in QCD finite when the cutoff is removed. This contact
term of the QCD Lagrangian is absorbed in the coupling constant 
$\beta_{12}(0)$ of the effective theory.
The renormalization of the coupling constant $\beta_{12}(0)$ thus involves the
renormalization factors relevant for $\beta $ and is not covered by the above
renormalization prescription of the potential $\beta_{12}$.
Since the contact term does not contribute here, we can safely neglect this
complication. 
Rescaling the singlet field as advocated in Sec. 3 modifies the potentials
according to
\begin{equation}
\begin{split}
        (\bar{\beta}_4 - i \bar{\beta}_{22})(x) = & (\beta_4 - i \beta_{22})({\scriptstyle \frac{F}{\sqrt{\lambda}}} x) 
                                                                e^{\frac{i}{3} (\sqrt{6}-\frac{F}{\sqrt{\lambda}}) x} \\
        (\bar{\beta}_5 - i \bar{\beta}_{21})(x) = & (\beta_5 - i \beta_{21})({\scriptstyle \frac{F}{\sqrt{\lambda}}} x) 
                                                                e^{\frac{i}{3} (\sqrt{6}-\frac{F}{\sqrt{\lambda}}) x} \\
        ( 2\bar{\beta}_5 + 3 \bar{\beta}_{18} -2 i \bar{\beta}_{21} +3 i \bar{\beta}_{23} )(x) = &
                        ( 2\beta_5 + 3 \beta_{18} -2 i \beta_{21} +3 i \beta_{23} )({\scriptstyle \frac{F}{\sqrt{\lambda}}} x)\\ 
                                        & \qquad \times e^{\frac{i}{3}
                                        (\sqrt{6}-\frac{F}{\sqrt{\lambda}}) x}
                                        {\textstyle \frac{F}{\sqrt{6 \lambda}}} \\
        ( 3 \bar{\beta}_4 + \bar{\beta}_5 -9 \bar{\beta}_{17} +3 \bar{\beta}_{18} -i \bar{\beta}_{21} \qquad & \\
                        -3 i \bar{\beta}_{22} +3 i \bar{\beta}_{23} -9 i \bar{\beta}_{24} )(x) =& 
                                ( 3 \beta_4 + \beta_5 -9 \beta_{17} +3 \beta_{18} -i \beta_{21} \\
                                & \qquad -3 i \beta_{22} +3 i \beta_{23} -9 i \beta_{24} )({\scriptstyle \frac{F}{\sqrt{\lambda}}} x) \\
                                        & \qquad \qquad \times e^{\frac{i}{3}
                                        (\sqrt{6}-\frac{F}{\sqrt{\lambda}}) x}
                                        {\textstyle \frac{F^2}{6 \lambda}} \\
        (\bar{\beta}_6 + \bar{\beta}_7 + i \bar{\beta}_{26})(x) = & (\beta_6 + \beta_7 + i \beta_{26})({\scriptstyle \frac{F}{\sqrt{\lambda}}} x) 
                                e^{\frac{2 i}{3} (\sqrt{6}-\frac{F}{\sqrt{\lambda}}) x} \\
        (\bar{\beta}_6 - \bar{\beta}_7)(x) = & (\beta_6 - \beta_7)({\scriptstyle \frac{F}{\sqrt{\lambda}}} x) \\
        \bar{\beta}_{12} (x) = & \beta_{12} ({\scriptstyle \frac{F}{\sqrt{\lambda}}} x) \\
        (\bar{\beta}_8+ i \bar{\beta}_{25})(x) = & (\beta_8 + i \beta_{25})({\scriptstyle \frac{F}{\sqrt{\lambda}}} x) 
                                e^{\frac{2 i}{3} (\sqrt{6}-\frac{F}{\sqrt{\lambda}}) x} \\
        (\bar{\beta}_{52} -i \bar{\beta}_{53})(x) = & (\beta_{52} -i \beta_{53})({\scriptstyle \frac{F}{\sqrt{\lambda}}} x) 
              e^{\frac{i}{3} (\sqrt{6}-\frac{F}{\sqrt{\lambda}}) x}
              {\textstyle \frac{F}{\sqrt{6 \lambda}}} .
\end{split}
\end{equation}
The $\bar{\beta}_i$ are the QCD scale invariant potentials which we use in our
calculation, and they are expanded in $\eta_0$
\begin{eqnarray}
\bar{\beta}_i(\frac{\eta_0}{F}) &=& \bar{\beta}_i^{(0)} + \bar{\beta}_i^{(2)} 
\frac{\eta_0^2}{F^2} +
\bar{\beta}_i^{(4)} \frac{\eta_0^4}{F^4} + \ldots
\q \mbox{for} \q i= 4, 5, 6, 7, 8, 12, 17,18, 53 \no \\
\bar{\beta}_i(\frac{\eta_0}{F}) &=& \bar{\beta}_i^{(1)} \frac{\eta_0}{F} + 
\bar{\beta}_i^{(3)} \frac{\eta_0^3}{F^3}  
+ \ldots \q \mbox{for} \q i= 21, 22, 23, 24, 25, 26, 52 .
\end{eqnarray}

\section{} \label{app:b}
\setcounter{equation}{0}
In this Appendix, we list the coefficients $C^{ab}_P, D^{ab}_P$ and
$\Delta_P$ from Eqs. (\ref{mass}) and (\ref{mass2}).
\begin{equation}
\begin{split}
        C_\pi^{\pi \pi} = & - 8 (\bar{\beta}_4^{(0)} + \bar{\beta}_5^{(0)} 
                                - 2 (\bar{\beta}_6^{(0)} + \bar{\beta}_8^{(0)})) \\
        C_\pi^{\pi K} = C_\pi^{K \pi} =& -8 (\bar{\beta}_4^{(0)} - 2 \bar{\beta}_6^{(0)}) \\
        C_K^{\pi K} = C_K^{K \pi} =& -4 (\bar{\beta}_4^{(0)} - 2 \bar{\beta}_6^{(0)}) \\
        C_K^{K K} = & - 8 ( 2 \bar{\beta}_4^{(0)} + \bar{\beta}_5^{(0)}
                                - 4 \bar{\beta}_6^{(0)} - 2 \bar{\beta}_8^{(0)}) \\
        C_\eta^{\pi \pi} = & - \frac{2}{3} ( 8 ( \bar{\beta}_6^{(0)} - 8 \bar{\beta}_7^{(0)} - 
                                                        3 \bar{\beta}_8^{(0)} ) \cos^2 \vartheta \\ 
                                & + 8 ( 2 \sqrt{2} (\bar{\beta}_6^{(0)} + \bar{\beta}_7^{(0)}) 
                                                        - \sqrt{3} \bar{\beta}_{26}^{(1)}) \sin (2 \vartheta) \\
                                & - (4( 2 \bar{\beta}_6^{(0)} - 3 \bar{\beta}_6^{(2)} + 2 \bar{\beta}_7^{(0)}
                                                + 6 \bar{\beta}_8^{(0)}) \\
                                        & \qquad - 9 ( 2 \bar{\beta}_8^{(2)} +
                                                \bar{\beta}_{12}^{(2)}) 
                                        -4 \sqrt{6}( 3 \bar{\beta}_{25}^{(1)} + \bar{\beta}_{26}^{(1)})) \sin^2 \vartheta )\\
        C_\eta^{\pi K} = C_\eta^{K \pi} =& \frac{4}{3} ( 4 ( \bar{\beta}_6^{(0)} - 4( 2 \bar{\beta}_7^{(0)} + 
                                                        \bar{\beta}_8^{(0)}) ) \cos^2 \vartheta \\
                                & - 2 ( 2 \sqrt{2} ( \bar{\beta}_6^{(0)} + \bar{\beta}_7^{(0)} +
                                                2 \bar{\beta}_8^{(0)}) - \sqrt{3} ( 2 \bar{\beta}_{25}^{(1)}
                                                + \bar{\beta}_{26}^{(1)}) ) \sin (2 \vartheta) \\
                                & + ( 4( 2 \bar{\beta}_6^{(0)} - 3 \bar{\beta}_6^{(2)} + 2 \bar{\beta}_7^{(0)}
                                                - 2 \bar{\beta}_8^{(0)}) \\
                                        & \qquad + 3 ( 2 \bar{\beta}_8^{(2)} + \bar{\beta}_{12}^{(2)}) 
                                                + 4 \sqrt{6} (\bar{\beta}_{25}^{(1)}
                                                        -\bar{\beta}_{26}^{(1)})) \sin^2 \vartheta ) \\
        C_\eta^{\pi \eta} = C_\eta^{\eta \pi } =& -\frac{2}{3} ( 
                                3 ( 2 \bar{\beta}_4^{(0)}- 3 \bar{\beta}_{17}^{(0)} + \bar{\beta}_{18}^{(0)})
                                - ( 2 \bar{\beta}_5^{(0)}- 9 \bar{\beta}_{17}^{(0)} + 3 \bar{\beta}_{18}^{(0)}) \cos ( 2 \vartheta ) \\
                                & -2 \sqrt{2} (2 \bar{\beta}_5^{(0)} 
                                            +3 \bar{\beta}_{18}^{(0)}) \sin (2 \vartheta)) \\
        C_\eta^{K K} =& \frac{8}{3} ( 16 (\bar{\beta}_6^{(0)}+\bar{\beta}_7^{(0)}
                                                +\bar{\beta}_8^{(0)}) \cos^2 \vartheta \\
                                & + 4 ( 2 \sqrt{2} (\bar{\beta}_6^{(0)}+\bar{\beta}_7^{(0)}
                                                +\bar{\beta}_8^{(0)}) -\sqrt{3} (\bar{\beta}_{25}^{(1)} 
                                                + \bar{\beta}_{26}^{(1)})) \sin (2 \vartheta) \\
                                & + ( 4 ( 2 \bar{\beta}_6^{(0)} - 3 \bar{\beta}_6^{(2)} + 2 \bar{\beta}_7^{(0)} 
                                                + 2 \bar{\beta}_8^{(0)}) \\
                                        & \qquad -3 ( 2 \bar{\beta}_8^{(2)} 
                                                + \bar{\beta}_{12}^{(2)}) - 4 \sqrt{6} ( \bar{\beta}_{25}^{(1)} 
                                                + \bar{\beta}_{26}^{(1)})) \sin^2 \vartheta) \\
        C_\eta^{K \eta} = C_\eta^{\eta K} =& -\frac{4}{3} ( 3 ( 2 \bar{\beta}_4^{(0)} + \bar{\beta}_5^{(0)} 
                                                        - 3 \bar{\beta}_{17}^{(0)} + \bar{\beta}_{18}^{(0)})
                                                + ( \bar{\beta}_5^{(0)} + 9 \bar{\beta}_{17}^{(0)} - 3 \bar{\beta}_{18}^{(0)}) 
                                                        \cos (2 \vartheta) \\
                                &+ \sqrt{2} (2 \bar{\beta}_5^{(0)} +3 \bar{\beta}_{18}^{(0)}) \sin (2 \vartheta)) \\
        C_{\eta'}^{\pi \pi} =& - \frac{2}{3} ( 8 ( \bar{\beta}_6^{(0)} - 8 \bar{\beta}_7^{(0)} - 
                                                        3 \bar{\beta}_8^{(0)} ) \sin^2 \vartheta \\ 
                                & - 8 ( 2 \sqrt{2} (\bar{\beta}_6^{(0)} + \bar{\beta}_7^{(0)}) 
                                                        - \sqrt{3} \bar{\beta}_{26}^{(1)}) \sin (2 \vartheta) \\
                                & - (4( 2 \bar{\beta}_6^{(0)} - 3 \bar{\beta}_6^{(2)} + 2 \bar{\beta}_7^{(0)}
                                                + 6 \bar{\beta}_8^{(0)}) \\
                                        & \qquad - 9 ( 2 \bar{\beta}_8^{(2)} +
                                                \bar{\beta}_{12}^{(2)}) 
                                        -4 \sqrt{6}( 3 \bar{\beta}_{25}^{(1)} + \bar{\beta}_{26}^{(1)})) \cos^2 \vartheta )\\
\end{split} \no
\end{equation}
\begin{equation}
\begin{split}
        C_{\eta'}^{\pi K} = C_{\eta'}^{K \pi} =& \frac{4}{3} ( 4 (\bar{\beta}_6^{(0)} -4( 2 \bar{\beta}_7^{(0)} 
                                                                        +\bar{\beta}_8^{(0)})) \sin^2 \vartheta \\
                                                &+2 (2 \sqrt{2} (\bar{\beta}_6^{(0)} +\bar{\beta}_7^{(0)} 
                                                                +2\bar{\beta}_8^{(0)}) -\sqrt{3} (2 \bar{\beta}_{25}^{(1)}
                                                                + \bar{\beta}_{26}^{(1)})) \sin (2 \vartheta) \\
                                                &+( 4 ( 2 \bar{\beta}_6^{(0)} -3 \bar{\beta}_6^{(2)}
                                                                +2 \bar{\beta}_7^{(0)} - 2 \bar{\beta}_8^{(0)})
                                                                +3( 2 \bar{\beta}_8^{(2)} + \bar{\beta}_{12}^{(2)})\\
                                                        & \qquad +4 \sqrt{6} (\bar{\beta}_{25}^{(1)} - \bar{\beta}_{26}^{(1)})) 
                                                                        \cos^2 \vartheta \\
        C_{\eta'}^{\pi \eta'} = C_{\eta'}^{\eta' \pi} =& -\frac{2}{3} ( 3 ( 2 \bar{\beta}_4^{(0)} - 3 \bar{\beta}_{17}^{(0)} 
                                                                        +\bar{\beta}_{18}^{(0)})
                                                + ( 2 \bar{\beta}_5^{(0)} - 9 \bar{\beta}_{17}^{(0)} + 3 \bar{\beta}_{18}^{(0)})
                                                        \cos ( 2 \vartheta ) \\
                                                &+ 2 \sqrt{2} (2 \bar{\beta}_5^{(0)} 
                                                                + 3 \bar{\beta}_{18}^{(0)}) \sin (2 \vartheta) )\\
        C_{\eta'}^{K K} =& \frac{8}{3} ( 16 (\bar{\beta}_6^{(0)}+\bar{\beta}_7^{(0)}
                                                +\bar{\beta}_8^{(0)}) \sin^2 \vartheta \\
                                & - 4 ( 2 \sqrt{2} (\bar{\beta}_6^{(0)}+\bar{\beta}_7^{(0)}
                                                +\bar{\beta}_8^{(0)}) -\sqrt{3} (\bar{\beta}_{25}^{(1)} 
                                                + \bar{\beta}_{26}^{(1)})) \sin (2 \vartheta) \\
                                & + ( 4 ( 2 \bar{\beta}_6^{(0)} - 3 \bar{\beta}_6^{(2)} + 2 \bar{\beta}_7^{(0)} 
                                                + 2 \bar{\beta}_8^{(0)}) \\
                                        & \qquad -3 ( 2 \bar{\beta}_8^{(2)} 
                                                + \bar{\beta}_{12}^{(2)}) - 4 \sqrt{6} ( \bar{\beta}_{25}^{(1)} 
                                                + \bar{\beta}_{26}^{(1)})) \cos^2 \vartheta)\\
        C_{\eta'}^{K \eta'} = C_{\eta'}^{\eta' K} =& -\frac{4}{3}( 3 ( 2 \bar{\beta}_4^{(0)} + \bar{\beta}_5^{(0)} 
                                                                -3 \bar{\beta}_{17}^{(0)} + \bar{\beta}_{18}^{(0)})
                                        - \sqrt{2} (2 \bar{\beta}_5^{(0)} + 3 \bar{\beta}_{18}^{(0)}) \sin (2 \vartheta) \\
                                        & - (\bar{\beta}_5^{(0)} + 9 \bar{\beta}_{17}^{(0)}
                                                - 3 \bar{\beta}_{18}^{(0)})
                                                  \cos (2 \vartheta) ) .
\end{split}
\end{equation} \\[1em]
\begin{equation}
\begin{split}
        D_\pi^{\pi \pi} =& \frac{1}{2} \\
        D_\pi^{\pi \eta} =& -\frac{1}{6 F^2} ( F^2 \cos^2 \vartheta - ( \sqrt{2} F^2 +4 \sqrt{3} \bar{v}_{3}^{(1)} ) \sin (2 \vartheta)\\
                                & + 2 (F^2 + 12 ( \bar{v}_{1}^{(2)} - \bar{v}_{2}^{(2)}) + 4 \sqrt{6} \bar{v}_{3}^{(1)}) \sin^2 \vartheta )\\
        D_\pi^{\pi \eta'} =& -\frac{1}{6 F^2} ( F^2 \sin^2 \vartheta + ( \sqrt{2} F^2 +4 \sqrt{3} \bar{v}_{3}^{(1)} ) \sin (2 \vartheta)\\
                                & + 2 (F^2 + 12 ( \bar{v}_{1}^{(2)} - \bar{v}_{2}^{(2)}) + 4 \sqrt{6} \bar{v}_{3}^{(1)}) \cos^2 \vartheta )\\
        D_K^{\pi \eta} =& \frac{\cos \vartheta}{12 F^2} ( F^2 \cos \vartheta + 2 ( \sqrt{2} F^2 
                                                                        +4 \sqrt{3} \bar{v}_{3}^{(1)}) \sin \vartheta) \\
        D_K^{\pi \eta'} =& \frac{\sin \vartheta}{12 F^2} ( F^2 \sin \vartheta - 2 ( \sqrt{2} F^2 
                                                                        +4 \sqrt{3} \bar{v}_{3}^{(1)}) \cos \vartheta) \\
        D_K^{K \eta} =&  -\frac{\sin \vartheta}{3 F^2} ( (F^2 + 4 ( 3 (\bar{v}_{1}^{(2)} -  \bar{v}_{2}^{(2)}) 
                                                                + \sqrt{6} \bar{v}_{3}^{(1)})) \sin \vartheta \\
                                                &+ ( \sqrt{2} F^2 +4 \sqrt{3} \bar{v}_{3}^{(1)}) \cos \vartheta) \\
        D_K^{K \eta'} =&  -\frac{\cos \vartheta}{3 F^2} ( (F^2 + 4 ( 3 (\bar{v}_{1}^{(2)} - \bar{v}_{2}^{(2)}) 
                                                                + \sqrt{6} \bar{v}_{3}^{(1)})) \cos \vartheta \\
                                                &-( \sqrt{2} F^2 +4 \sqrt{3} \bar{v}_{3}^{(1)}) \sin \vartheta)
\end{split} \no
\end{equation}
\begin{equation}
\begin{split}
        D_K^{\eta \eta} =& \frac{\cos^2 \vartheta}{4} \\
        D_K^{\eta' \eta'} =& \frac{\sin^2 \vartheta}{4} \\
        D_\eta^{\pi \pi} =& -\frac{1}{2 F^2} ( F^2 \cos^2 \vartheta - (\sqrt{2} F^2 + 4 \sqrt{3} \bar{v}_{3}^{(1)}) \sin (2 \vartheta) \\
                                                &+2 (F^2 + 12 (\bar{v}_{1}^{(2)} - \bar{v}_{2}^{(2)}) 
                                                        + 4 \sqrt{6} \bar{v}_{3}^{(1)}) \sin^2 \vartheta ) \\
        D_\eta^{\pi K} =& \frac{\cos \vartheta}{3 F^2} ( F^2 \cos \vartheta + 2 ( \sqrt{2} F^2 
                                                                        +4 \sqrt{3} \bar{v}_{3}^{(1)}) \sin \vartheta) \\
        D_\eta^{\pi \eta} =& \frac{1}{18 F^2} ( 7 F^2 \cos^4 \vartheta
                                        + 2 (5 \sqrt{2} F^2 + 4 \sqrt{3} \bar{v}_{3}^{(1)})  \cos^2 \vartheta \sin (2 \vartheta) \\
                                        & + 3 (F^2 +4 \sqrt{6} \bar{v}_{3}^{(1)} -12 \bar{v}_{2}^{(2)}) \sin^2 (2 \vartheta) \\
                                        & + 8 (\sqrt{2} (F^2 -36\bar{v}_{2}^{(2)}) + 12 \sqrt{3} (\bar{v}_{3}^{(1)}
                                                                - 3 \bar{v}_{3}^{(3)})) \sin^2 \vartheta \sin (2\vartheta) \\
                                        & - 2 (F^2 - 72( \bar{v}_{2}^{(2)} - 3 \bar{v}_{2}^{(4)})
                                                                +8 \sqrt{6} ( \bar{v}_{3}^{(1)} - 9 \bar{v}_{3}^{(3)})) \sin^4 \vartheta)\\
        D_\eta^{\pi \eta'} = D_{\eta'}^{\pi \eta} =& \frac{1}{144 F^2} ( 3 (3 F^2 +32 \bar{v}_{2}^{(2)} -144 \bar{v}_{2}^{(4)}
                                                + 48 \sqrt{6} \bar{v}_{3}^{(3)}) \\
                                &+ (7 F^2 -16( 18 \bar{v}_{2}^{(2)} -27 \bar{v}_{2}^{(4)}
                                                - \sqrt{6}( 4 \bar{v}_{3}^{(1)} - 9 \bar{v}_{3}^{(3)}))) \cos (4 \vartheta) \\
                                &- 4 ( \sqrt{2} ( F^2 + 144 \bar{v}_{2}^{(2)}) - 4 \sqrt{3}( 7 \bar{v}_{3}^{(1)}
                                                - 36 \bar{v}_{3}^{(3)})) \sin (4 \vartheta)) \\
        D_\eta^{K K} =&  -\frac{4 \sin \vartheta}{3 F^2} ( ( F^2 + 4( 3 (\bar{v}_{1}^{(2)} - \bar{v}_{2}^{(2)}) 
                                                                +\sqrt{6} \bar{v}_{3}^{(1)})) \sin \vartheta \\
                                                &+ (\sqrt{2} F^2 +4 \sqrt{3} \bar{v}_{3}^{(1)}) \cos \vartheta)\\
        D_\eta^{K \eta} =& -\frac{2}{9 F^2} ( 4 F^2 \cos^4 \vartheta 
                                        + 4 (\sqrt{2} F^2 + 4 \sqrt{3} \bar{v}_{3}^{(1)}) \cos^2 \vartheta \sin (2 \vartheta) \\
                                        &+ 3 (F^2 + 4 \sqrt{6} \bar{v}_{3}^{(1)} - 12 \bar{v}_{2}^{(2)}) \sin^2 (2 \vartheta) \\
                                        &+ 2 ( \sqrt{2} (F^2 -36 \bar{v}_{2}^{(2)}) + 12 \sqrt{3} (\bar{v}_{3}^{(1)} 
                                                                -3 \bar{v}_{3}^{(3)})) \sin^2 \vartheta \sin (2 \vartheta) \\
                                        &+ (F^2 -8 ( 9 (\bar{v}_{2}^{(2)} - 3 \bar{v}_{2}^{(4)})
                                                                - \sqrt{6} (\bar{v}_{3}^{(1)} - 9 \bar{v}_{3}^{(3)}))) \sin^4 \vartheta) \\
        D_\eta^{K \eta'} = D_{\eta'}^{K \eta} =& -\frac{1}{36 F^2} ( 3 ( 3 F^2 - 8 ( 5 \bar{v}_{2}^{(2)} -9 \bar{v}_{2}^{(4)}
                                                - \sqrt{6} ( \bar{v}_{3}^{(1)} - 3 \bar{v}_{3}^{(3)}))) \\
                                &+ (7 F^2 - 8 ( 9 \bar{v}_{2}^{(2)} +27 \bar{v}_{2}^{(4)}
                                                - \sqrt{6} ( 5 \bar{v}_{3}^{(1)} +9 \bar{v}_{3}^{(3)}))) \cos (4 \vartheta) \\
                                &- 4 ( \sqrt{2} (F^2 + 36 \bar{v}_{2}^{(2)}) - 4 \sqrt{3} (\bar{v}_{3}^{(1)} 
                                                - 9 \bar{v}_{3}^{(3)})) \sin (4 \vartheta))\\
        D_\eta^{\eta K} =& \cos^2 \vartheta \\
        D_\eta^{\eta \eta} =& -\frac{4 \sin^2 \vartheta}{F^2} ( 2 \bar{v}_{1}^{(2)} - 3 \bar{v}_{4}^{(2)}
                                        + 3 \bar{v}_{4}^{(2)} \cos (2 \vartheta) ) \\
        D_\eta^{\eta \eta'} = D_{\eta'}^{\eta \eta} =& -\frac{2 \cos^2 \vartheta}{F^2}  ( 2 \bar{v}_{1}^{(2)} - 3 \bar{v}_{4}^{(2)}
                                        + 3 \bar{v}_{4}^{(2)} \cos (2 \vartheta) ) \\
        D_\eta^{\eta' \eta'} = D_{\eta'}^{\eta' \eta} =& -\frac{2 \sin^2 \vartheta}{F^2} ( 2 \bar{v}_{1}^{(2)} - 3 \bar{v}_{4}^{(2)}
                                        - 3 \bar{v}_{4}^{(2)} \cos (2 \vartheta) )
\end{split} \no
\end{equation}
\begin{equation}
\begin{split}
        D_{\eta'}^{\pi \pi} =& -\frac{1}{2 F^2} ( F^2 \sin^2 \vartheta + ( \sqrt{2} F^2 + 4 \sqrt{3} \bar{v}_{3}^{(1)}) \sin (2 \vartheta) \\
                                                &+2 (F^2 + 4 (3\bar{v}_{1}^{(2)} - 3 \bar{v}_{2}^{(2)} 
                                                                + \sqrt{6} \bar{v}_{3}^{(1)})) \cos^2 \vartheta ) \\
        D_{\eta'}^{\pi K} =& \frac{\sin \vartheta}{3 F^2} ( F^2 \sin \vartheta  
                                                        - 2 ( \sqrt{2} F^2 +4 \sqrt{3} \bar{v}_{3}^{(1)}) \cos \vartheta) \\
        D_{\eta'}^{\pi \eta'} =& \frac{1}{18 F^2} ( 7 F^2 \sin^4 \vartheta
                                        - 2 (5 \sqrt{2} F^2 + 4 \sqrt{3} \bar{v}_{3}^{(1)})  \sin^2 \vartheta \sin (2 \vartheta) \\
                                        & + 3 (F^2 +4 \sqrt{3} \bar{v}_{3}^{(1)} -12 \bar{v}_{2}^{(2)}) \sin^2 (2 \vartheta) \\
                                        & - 8 ( \sqrt{2} (F^2 -36 \bar{v}_{2}^{(2)}) + 12 \sqrt{3} ( \bar{v}_{3}^{(1)}
                                                                +3 \bar{v}_{3}^{(3)})) \cos^2 \vartheta \sin (2\vartheta) \\
                                        & - 2 (F^2 - 8 ( 9 \bar{v}_{2}^{(2)} - 27 \bar{v}_{2}^{(4)}
                                                                - \sqrt{6} ( \bar{v}_{3}^{(1)} -9 \bar{v}_{3}^{(3)}))) \cos^4 \vartheta)\\
        D_{\eta'}^{K K} =& -\frac{4 \cos \vartheta}{3 F^2} ( (F^2 + 4 ( 3 \bar{v}_{1}^{(2)} - 3 \bar{v}_{2}^{(2)} 
                                                                + \sqrt{6} \bar{v}_{3}^{(1)})) \cos \vartheta \\
                                                &- ( \sqrt{2} F^2 +4 \sqrt{3} \bar{v}_{3}^{(1)}) \sin \vartheta) \\
        D_{\eta'}^{K \eta'} =& -\frac{1}{9 F^2} ( 8 F^2 \sin^4 \vartheta 
                                        - 8 (\sqrt{2} F^2 + 4 \sqrt{3} \bar{v}_{3}^{(1)}) \sin^2 \vartheta \sin (2 \vartheta) \\
                                        &+ 6 (F^2 + 4 \sqrt{6} \bar{v}_{3}^{(1)} - 12 \bar{v}_{2}^{(2)}) \sin^2 (2 \vartheta) \\
                                        &- 4 ( \sqrt{2} (F^2 -36 \bar{v}_{2}^{(2)}) + 12 \sqrt{3}( \bar{v}_{3}^{(1)} 
                                                                +3 \bar{v}_{3}^{(3)})) \cos^2 \vartheta \sin (2 \vartheta) \\
                                        &+ 2 (F^2 -8 ( 9 \bar{v}_{2}^{(2)} - 27 \bar{v}_{2}^{(4)}
                                                                - \sqrt{6} ( \bar{v}_{3}^{(1)} - 9 \bar{v}_{3}^{(3)}))) \cos^4 \vartheta )\\
        D_{\eta'}^{\eta' K} =& \sin^2 \vartheta\\
        D_{\eta'}^{\eta' \eta'} =&  -\frac{4 \cos^2 \vartheta}{F^2} (2 \bar{v}_{1}^{(2)} - 3 \bar{v}_{4}^{(2)} 
                                        -3 \bar{v}_{4}^{(2)} \cos (2 \vartheta) ) .
\end{split}
\end{equation} \\[1em]
\begin{equation}
\begin{split}
        \Delta_\eta =& (\overcirc{m}^2_{\eta'}-\overcirc{m}^2_\eta) \sin^2 \vartheta
                        + (\overcirc{m}^2_K - \overcirc{m}^2_\pi )\frac{2 ( \sqrt{2} F^2 + 4 \sqrt{3} \bar{v}_3^{(1)})}{3 F^2} 
                                \sin (2 \vartheta) \\
                        & + 12 \bar{v}_0^{(4)} \sin^4 \vartheta \frac{1}{16 \pi^2 F^4} m_\eta^2
                                        \ln \frac{m_\eta^2}{\mu^2}\\
        \Delta_{\eta'} =& (\overcirc{m}^2_\eta-\overcirc{m}^2_{\eta'}) \sin^2 \vartheta
                        - (\overcirc{m}^2_K - \overcirc{m}^2_\pi )\frac{2 ( \sqrt{2} F^2 + 4 \sqrt{3} \bar{v}_3^{(1)})}{3 F^2} 
                                \sin (2 \vartheta) \\
                        & + 3 \bar{v}_0^{(4)} \sin^2 (2 \vartheta) \frac{1}{16 \pi^2 F^4} 
                                m_\eta^2 \ln \frac{m_\eta^2}{\mu^2}\\
\Delta_\eta^{(dim)} =& 3 \bar{v}_0^{(4)} \sin^2 ( 2 \vartheta) 
         \frac{1}{16 \pi^2 F^4} m_{\eta'}^2\ln \frac{m_{\eta'}^2}{\mu^2}\\
\Delta_{\eta'}^{(dim)} =& 12 \bar{v}_0^{(4)} \cos^4 \vartheta 
         \frac{1}{16 \pi^2 F^4} m_{\eta'}^2 \ln \frac{m_{\eta'}^2}{\mu^2}\\
\end{split}
\end{equation}

\section{} \label{app:c}
\setcounter{equation}{0}
The coefficients $G^a_\phi$ and  $H^a_\phi$ of 
Eqs. (\ref{dec}) and (\ref{dec2}) read 
\begin{equation}
\begin{split}
        G_\pi^\pi =& 4 (\bar{\beta}_4^{(0)} + \bar{\beta}_5^{(0)}) \\
        G_\pi^K =& 8 \bar{\beta}_4^{(0)} \\
        G_K^\pi =& 4 \bar{\beta}_4^{(0)} \\
        G_K^K =&  8 \bar{\beta}_4^{(0)} + 4 \bar{\beta}_4^{(0)} \\
        G_8^\pi =& \frac{1}{3} ( 3 ( 4 \bar{\beta}_4^{(0)} - 2 \bar{\beta}_5^{(0)} 
                                        + 3 \bar{\beta}_{17}^{(0)} - \bar{\beta}_{18}^{(0)}) \\
                                &+ (2 \bar{\beta}_5^{(0)} 
                                        - 9 \bar{\beta}_{17}^{(0)} + 3 \bar{\beta}_{18}^{(0)}) \cos (4 \vartheta) \\
                                &+ 2 \sqrt{2} (2 \bar{\beta}_5^{(0)} + 3 \bar{\beta}_{18}^{(0)}) \sin (4 \vartheta) ) \\
        G_8^K =& \frac{2}{3} ( 3 ( 4 \bar{\beta}_4^{(0)}+ 3 \bar{\beta}_5^{(0)} 
                                        + 3 \bar{\beta}_{17}^{(0)} - \bar{\beta}_{18}^{(0)}) \\
                                &- ( \bar{\beta}_5^{(0)} 
                                        + 9 \bar{\beta}_{17}^{(0)} - 3 \bar{\beta}_{18}^{(0)}) \cos (4 \vartheta) \\
                                &- \sqrt{2} (2 \bar{\beta}_5^{(0)} + 3 \bar{\beta}_{18}^{(0)}) \sin (4 \vartheta) ) \\
        G_0^\pi =& \frac{1}{3} ((12 \bar{\beta}_4^{(0)} + 6 \bar{\beta}_5^{(0)} -45 \bar{\beta}_{17}^{(0)}
                                                                + 15 \bar{\beta}_{18}^{(0)} - 6 \sqrt{6} \bar{\beta}_{52}^{(1)}) \\
                                & - ( 2 \bar{\beta}_5^{(0)} - 9 \bar{\beta}_{17}^{(0)}
                                                                +3 \bar{\beta}_{18}^{(0)}) \cos (4 \vartheta) \\
                                & - 2 \sqrt{2} ( 2 \bar{\beta}_5^{(0)} + 3 \bar{\beta}_{18}^{(0)}) \sin (4 \vartheta)) \\
        G_0^K =& \frac{2}{3} ( 3 ( 4 \bar{\beta}_4^{(0)} + \bar{\beta}_5^{(0)} -15 \bar{\beta}_{17}^{(0)}
                                                                + 5 \bar{\beta}_{18}^{(0)} -2 \sqrt{6} \bar{\beta}_{52}^{(1)}) \\
                                & + ( \bar{\beta}_5^{(0)} + 9 \bar{\beta}_{17}^{(0)}
                                                                -3 \bar{\beta}_{18}^{(0)}) \cos (4 \vartheta) \\
                                & + \sqrt{2} ( 2 \bar{\beta}_5^{(0)} + 3
                                \bar{\beta}_{18}^{(0)}) \sin (4 \vartheta)) .
\end{split}
\end{equation}
\begin{equation}
\begin{split}
        H_\pi^\pi =& - 1 \\
        H_\pi^K =& - \frac{1}{2} \\
        H_\pi^\eta =& \frac{2}{F^2} \bar{v}_1^{(2)} \sin^2 \vartheta \\
        H_\pi^{\eta'} =& \frac{2}{F^2} \bar{v}_1^{(2)} \cos^2 \vartheta \\
        H_K^\pi =& - \frac{3}{8} \\
        H_K^K =& - \frac{3}{4} \\
        H_K^\eta =& - \frac{1}{8 F^2} ( 3 F^2 \cos^2 \vartheta - 16 \bar{v}_1^{(2)} \sin^2 \vartheta) \\
        H_K^{\eta'} =& - \frac{1}{8 F^2} ( 3 F^2 \sin^2 \vartheta - 16 \bar{v}_1^{(2)} \cos^2 \vartheta) \\
        H_8^K =& - \frac{1}{8} ( 13 - \cos (4 \vartheta) ) \\
        H_8^\eta =& \frac{\sin^2 \vartheta}{2 F^2} ( 4 \bar{v}_1^{(2)} + 3 \bar{v}_4^{(2)} 
                                        - 3 \bar{v}_4^{(2)} \cos (4 \vartheta)) \\
        H_8^{\eta'} =& \frac{\cos^2 \vartheta}{2 F^2} ( 4 \bar{v}_1^{(2)} + 3 \bar{v}_4^{(2)} 
                                        - 3 \bar{v}_4^{(2)} \cos (4 \vartheta)) \\
        H_0^K =& \frac{1}{4} \sin^2 (2 \vartheta) \\
        H_0^\eta =&  \frac{\sin^2 \vartheta}{2 F^2} ( 4 \bar{v}_1^{(2)} - 15 \bar{v}_4^{(2)}
                                + 3 \bar{v}_4^{(2)} \cos (4 \vartheta)) \\
        H_0^{\eta'} =&  \frac{\cos^2 \vartheta}{2 F^2} ( 4 \bar{v}_1^{(2)} - 15 \bar{v}_4^{(2)}
                                + 3 \bar{v}_4^{(2)} \cos (4 \vartheta)) .
\end{split}
\end{equation} \\
The $Z$-factors are given by
\begin{equation}
Z_P = 1 + \frac{1}{F^2} K^{a}_P  m_a^2  +
\frac{1}{16 \pi ^2 F^2} L^{a}_P   m_a^2 \ln \frac{m_a^2}{\mu^2}
\end{equation}
with the coefficients
\begin{equation}
\begin{split}
        K_\pi^\pi =& - 8 (\bar{\beta}_4^{(0)} +  \bar{\beta}_5^{(0)}) 
                \\
        K_\pi^K =& -16 \bar{\beta}_4^{(0)} 
                \\
        K_K^\pi =& -8 \bar{\beta}_4^{(0)} 
                \\
        K_K^K =&-8 ( 2 \bar{\beta}_4^{(0)} - \bar{\beta}_5^{(0)})
                \\
        K_\eta^\pi =& -\frac{4}{3} (  6 \bar{\beta}_4^{(0)} - 9 \bar{\beta}_{17}^{(0)} +
                                        3 \bar{\beta}_{18}^{(0)}
                        -(2 \bar{\beta}_5^{(0)} - 9 \bar{\beta}_{17}^{(0)} + 3 \bar{\beta}_{18}^{(0)}) 
                                \cos (2 \vartheta) \\
                        &- 2 \sqrt{2} ( 2 \bar{\beta}_5^{(0)} + 3 \bar{\beta}_{18}^{(0)} ) \sin (2\vartheta))
                \\
        K_\eta^K =& -\frac{8}{3} (  3 ( 2 \bar{\beta}_4^{(0)} + \bar{\beta}_5^{(0)} - 3 \bar{\beta}_{17}^{(0)} +
                                         \bar{\beta}_{18}^{(0)})
                        +( \bar{\beta}_5^{(0)} + 9 \bar{\beta}_{17}^{(0)} - 3 \bar{\beta}_{18}^{(0)}) 
                                \cos (2 \vartheta) \\
                        &+\sqrt{2} ( 2 \bar{\beta}_5^{(0)} + 3 \bar{\beta}_{18}^{(0)} ) \sin (2\vartheta)) 
                \\
        K_{\eta'}^\pi =&-\frac{4}{3} (  6 \bar{\beta}_4^{(0)} - 9 \bar{\beta}_{17}^{(0)} +
                                        3 \bar{\beta}_{18}^{(0)}
                        +(2 \bar{\beta}_5^{(0)} - 9 \bar{\beta}_{17}^{(0)} + 3 \bar{\beta}_{18}^{(0)}) 
                                \cos (2 \vartheta) \\
                        &+ 2 \sqrt{2} ( 2 \bar{\beta}_5^{(0)} + 3 \bar{\beta}_{18}^{(0)} ) \sin (2\vartheta))
                \\
        K_{\eta'}^K =& -\frac{8}{3} (  3 ( 2 \bar{\beta}_4^{(0)} + \bar{\beta}_5^{(0)} - 3 \bar{\beta}_{17}^{(0)} +
                                         \bar{\beta}_{18}^{(0)})
                        -( \bar{\beta}_5^{(0)} + 9 \bar{\beta}_{17}^{(0)} - 3 \bar{\beta}_{18}^{(0)}) 
                                \cos (2 \vartheta) \\
                        &-\sqrt{2} ( 2 \bar{\beta}_5^{(0)} + 3 \bar{\beta}_{18}^{(0)} ) \sin (2\vartheta)). 
\end{split}
\end{equation}
\begin{equation}
\begin{split}
        L_\pi^\pi =& \frac{2}{3} \\
        L_\pi^K =& \frac{1}{3} \\
        L_\pi^\eta =& - \frac{4}{F^2} \bar{v}_1^{(2)} \sin^2 \vartheta \\
        L_\pi^{\eta'} =& - \frac{4}{F^2} \bar{v}_1^{(2)} \cos^2 \vartheta \\
        L_K^\pi =& \frac{1}{4} \\
        L_K^K =& \frac{1}{2} \\
        L_K^\eta =& \frac{1}{4 F^2} ( F^2 \cos^2 \vartheta- 16 \bar{v}_1^{(2)} \sin^2 \vartheta) \\
        L_K^{\eta'} =& \frac{1}{4 F^2} ( F^2 \sin^2 \vartheta- 16 \bar{v}_1^{(2)} \cos^2 \vartheta) \\
        L_\eta^K =& \cos^2 \vartheta \\
        L_\eta^\eta =& - \frac{2 \sin^2 \vartheta}{F^2} (2 \bar{v}_1^{(2)} - 3 \bar{v}_4^{(2)} 
                                                + 3 \bar{v}_4^{(2)} \cos (2 \vartheta)) \\
        L_\eta^{\eta'} =& - \frac{2 \cos^2 \vartheta}{F^2} (2 \bar{v}_1^{(2)} - 3 \bar{v}_4^{(2)} 
                                                + 3 \bar{v}_4^{(2)} \cos (2 \vartheta)) \\
        L_{\eta'}^K =& \sin^2 \vartheta \\
        L_{\eta'}^\eta =& - \frac{2 \sin^2 \vartheta}{F^2} (2 \bar{v}_1^{(2)} - 3 \bar{v}_4^{(2)} 
                                                - 3 \bar{v}_4^{(2)} \cos (2 \vartheta)) \\
        L_{\eta'}^{\eta'} =& - \frac{2 \cos^2 \vartheta}{F^2} (2 \bar{v}_1^{(2)} - 3 \bar{v}_4^{(2)} 
                - 3 \bar{v}_4^{(2)} \cos (2 \vartheta)) .
\end{split}
\end{equation}

\newpage

\section*{Table captions}

\begin{enumerate}

\item[Table 1] Shown are the next-to-leading order mass contributions in 
units of 10$^{-3}$ GeV$^2$ both in dimensional and infrared regularization. 
For the scale $\mu$ of the chiral logarithms we used  $\mu =m_{\rho}$, 
$\mu = m_{\eta'}$ and $\mu = \Lambda_\chi$ and 
employed the mixing angles $\vartheta = 0^\circ$ and 
$\vartheta = -20^\circ$. Only the couplings from standard chiral perturbation
theory have been retained while neglecting the remaining LECs.

\item[Table 2] The dependence of the next-to-leading order mass contributions
on the unknown parameters $\bar{v}_i^{(j)}$ is given in units of GeV$^2$. 
We used $\mu = m_{\rho}$ for the scale of the chiral logarithms and an
$\eta$-$\eta'$ mixing angle of $\vartheta= -20^\circ$.

\item[Table 3] Given are the next-to-leading order contributions to the
pseudoscalar decay constants in units of GeV both in infrared and in 
dimensional
regularization. The first column of each regularization scheme shows the
contribution which arises if only known LECs from standard chiral perturbation
theory are kept whereas the second and third columns show the dependence 
on $\bar{v}_1^{(2)}$ and $\bar{v}_4^{(2)}$.
We used $\mu = m_{\rho}$ for the scale of the chiral logarithms and an
$\eta$-$\eta'$ mixing angle of $\vartheta= -20^\circ$.

\end{enumerate}

\newpage

\begin{center}

\begin{small}
\begin{tabular}{|c|c|c|c|c|c|c|c|c|c|}
        \hline
        \multicolumn{2}{|c|}{\multirow{2}{3.9em}{$m^2_P - \overcirc{m}^2_P$}} 
        & \multicolumn{4}{|c|}{$\vartheta = 0^\circ$} & \multicolumn{4}{|c|}{$\vartheta = -20^\circ$} \\ \cline{3-10}
        \multicolumn{2}{|c|}{}  & $\pi$ & $K$ & $\eta$ & $\eta'$ & $\pi$ & $K$ & $\eta$ & $\eta'$ \\ \hline \hline
        \multirow{3}{0.7em}{\begin{sideways} ir \end{sideways}} 
        & \rule{0em}{1.4em} $\mu=m_\rho$ & -0.79 & -0.88 & -13.52 & 23.53 & -0.27 & -3.43 & -91.93 & -56.01 \\ \cline{2-10}
        & \rule{0em}{1.4em} $\mu=m_{\eta'}$ & -0.52 & -8.19 & -17.30 & 59.23 & 0.31 & -12.38 & -115.84 & -20.84 \\ \cline{2-10}
        & \rule{0em}{1.4em} $\mu=\Lambda_\chi$ & -0.26 & -15.74 & -21.19 & 96.03 & 0.90 & -21.60 & -140.49 & 15.42 \\ \hline \hline
        \multirow{3}{0.7em}{\begin{sideways} dim \end{sideways}} 
        & \rule{0em}{1.4em} $\mu=m_\rho$ & -2.69 & -24.76 & -44.73 & 6.97 & -1.20 & -27.00 & -122.72 & -119.04 \\ \cline{2-10}
        & \rule{0em}{1.4em} $\mu=m_{\eta'}$ & -0.52 & -8.19 & -17.30 & 59.23 & 0.31 & -12.38 & -115.84 & -20.84 \\ \cline{2-10}
        & \rule{0em}{1.4em} $\mu=\Lambda_\chi$ & 1.72 & 8.89 & 10.98 & 113.11 & 1.86 & 2.70 & -108.74 & 80.41 \\ \hline
\end{tabular}
\end{small}

\vfill

Table 1

\vfill
\vfill


\begin{tabular}{|l|c|c|c|c|c|c|c|c|}
        \hline
        \multirow{2}{3em}{} & \multicolumn{2}{|c|}{\rule{0em}{1.4em} $m^2_\pi - \overcirc{m}^2_\pi$} 
        & \multicolumn{2}{|c|}{$m^2_K - \overcirc{m}^2_K$} 
        & \multicolumn{2}{|c|}{$m^2_\eta - \overcirc{m}^2_\eta$} 
        & \multicolumn{2}{|c|}{$m^2_{\eta'} - \overcirc{m}^2_{\eta'}$} \\ \cline{2-9}
        & ir & dim & ir & dim & ir & dim & ir & dim \\ \hline \hline
        \rule{0em}{1.2em} $\propto \bar{v}_0^{(4)}$ & 0 & 0 & 0 & 0 & -2.84 & 39.23 & -21.47 & 296.14 \\ \hline
        \rule{0em}{1.2em} $\propto \bar{v}_1^{(2)}$ & 0.16 & -2.19 & 1.98 & -27.28 & 13.38 & -37.07 & 90.15 & -129.92 \\ \hline
        \rule{0em}{1.2em} $\propto \bar{v}_2^{(2)}$ & -0.16 & 2.19 & -1.98 & 27.28 & -16.57 & -28.31 & -58.41 & 173.76 \\ \hline
        \rule{0em}{1.2em} $\propto \bar{v}_2^{(4)}$ & 0 & 0 & 0 & 0 & 2.89 & 39.83 & 21.80 & -300.69 \\ \hline
        \rule{0em}{1.2em} $\propto \bar{v}_3^{(1)}$ & 0.38 & -1.04 & -1.40 & -31.19 & -88.47 & -100.39 & 148.22 & 53.31 \\ \hline
        \rule{0em}{1.2em} $\propto \bar{v}_3^{(3)}$ & 0 & 0 & 0 & 0 & 5.74 & -11.36 & 8.73 & 152.20 \\ \hline
        \rule{0em}{1.2em} $\propto \bar{v}_4^{(2)}$ & 0 & 0 & 0 & 0 & -1.70 & 49.50 & -26.13 & 556.86 \\ \hline
\end{tabular}

\vfill

Table 2

\vfill
\vfill


\begin{tabular}{|l|c|c|c|c|c|c|}
        \hline
        \multirow{2}{3em}{} & \multicolumn{3}{|c|}{\rule{0em}{1.2em} infrared} 
        & \multicolumn{3}{|c|}{dimensional}\\ \cline{2-7}
        & \rule{0em}{1.2em} $\propto 1$ & $\propto \bar{v}_1^{(2)}$ & $\propto \bar{v}_4^{(2)}$ 
        & \rule{0em}{1.2em} $\propto 1$ & $\propto \bar{v}_1^{(2)}$ & $\propto \bar{v}_4^{(2)}$ \\ \hline \hline
        \rule{0em}{1.1em} $F_\pi - F$ & 0.0066 & -0.3769 & 0 & 0.0066 & 5.1983 & 0 \\ \hline
        \rule{0em}{1.1em} $F_K - F$ & 0.0255 & -0.3769 & 0 & 0.0243 & 5.1983 & 0 \\ \hline
        \rule{0em}{1.1em} $F_8 - F$ & 0.0444 & -0.3769 & -0.2336 & 0.0444 & 5.1983 & 3.2217 \\ \hline
        \rule{0em}{1.1em} $\bar{F}_0 - F$ & -0.0060 & -0.3769 & 1.3643 & -0.0060 & 5.1983 & -18.8266 \\ \hline
\end{tabular}

\vfill

Table 3

\end{center}

\end{document}